\documentclass[useAMS,usenatbib,letterpaper,10pt]{mn2e}
\usepackage{times}
\usepackage{epsfig}
\usepackage{natbib}
\usepackage{amssymb}
\usepackage{amsmath}
\usepackage[bookmarks,bookmarksnumbered,colorlinks=true, citecolor=blue, linkcolor=black]{hyperref}
\usepackage{color}
\usepackage[usenames,dvipsnames]{xcolor} 

\usepackage[normalem]{ulem}

\newcommand{\Msun}{M$_\odot$} 

\title[The diverse evolutionary paths of simulated high-z massive, compact galaxies to z=0]{The diverse evolutionary paths of simulated high-z massive, compact galaxies to z=0}
\author[S. Wellons et al.]{Sarah Wellons$^{1}$\thanks{E-mail: swellons@cfa.harvard.edu}, Paul Torrey$^{2,3}$, Chung-Pei Ma$^{4}$, Vicente Rodriguez-Gomez$^{1}$, 
\newauthor Annalisa Pillepich$^{1}$, Dylan Nelson$^{1}$, Shy Genel$^{5}$\thanks{Hubble fellow}, Mark Vogelsberger$^{2}$, and Lars Hernquist$^{1}$ \\
$^{1}$Harvard-Smithsonian Center for Astrophysics, 60 Garden St., Cambridge, MA 02138, USA\\
$^{2}$Massachusetts Institute of Technology, Cambridge, MA 02139, USA \\
$^{3}$TAPIR, Mailcode 350-17, California Institute of Technology, Pasadena, CA 91125, USA \\
$^{4}$University of California Berkeley, Berkeley, CA 94720, USA \\
$^{5}$Department of Astronomy, Columbia University, 550 West 120th Street, New York, NY, 10027, USA}
\begin{document}

\maketitle

\label{firstpage}

\begin{abstract}
Massive quiescent galaxies have much smaller physical sizes at high redshift than today.  The strong evolution of galaxy size may be caused by progenitor bias, major and minor mergers, adiabatic expansion, and/or renewed star formation, but it is difficult to test these theories observationally.  Herein, we select a sample of 35 massive, compact galaxies ($M_* = 1-3 \times 10^{11}$ \Msun, $M_*/R^{1.5} > 10^{10.5}$ \Msun/kpc$^{1.5}$) at $z=2$ in the cosmological hydrodynamical simulation Illustris and trace them forward to $z=0$ to uncover their evolution and identify their descendants.  By $z=0$, the original factor of 3 difference in stellar mass spreads to a factor of 20.  The dark matter halo masses similarly spread from a factor of 5 to 40.  The galaxies' evolutionary paths are diverse: about half acquire an ex-situ envelope and are the core of a more massive descendant, a third survive undisturbed and gain very little mass, 15\% are consumed in a merger with a more massive galaxy, and a small remainder are thoroughly mixed by major mergers.  The galaxies grow in size as well as mass, and only $\sim$10\% remain compact by $z=0$.  The majority of the size growth is driven by the acquisition of ex-situ mass.  The most massive galaxies at $z=0$ are the most likely to have compact progenitors, but this trend possesses significant dispersion which precludes a direct linkage to compact galaxies at $z=2$.  The compact galaxies' merger rates are influenced by their $z=2$ environments, so that isolated or satellite compact galaxies (which are protected from mergers) are the most likely to survive to the present day.
\end{abstract}


\section{Introduction}
\label{sec:intro}

Massive galaxies at high redshift have been observed to have properties which can be quite different from their counterparts at similar mass in the local universe.  One of the most striking differences is the much smaller sizes observed for high-redshift galaxies.  For example, a quiescent galaxy with a stellar mass of $10^{11}$ M$_\odot$, which has a typical effective radius of 8 kpc in the local universe, has a typical size of only 1.7 kpc at $z=2$ \citep{VanderWel2014}.  Early reports of these extremely small sizes based on their spatial extent \citep{Daddi2005, VanDokkum2006, Trujillo2006a} have been confirmed by dynamical measurements of their stars \citep{VanDokkum2009, Cappellari2009, Newman2010, Belli2014} and, more recently, gas \citep{Barro2014a, Nelson2014}, in addition to  higher-resolution HST imaging \citep{VanDokkum2008, Szomoru2012, Williams2014}.  Several observational studies have tracked the strong evolution of galaxy size with redshift \citep{Trujillo2007, Franx2008, Taylor2010, VanderWel2014}, which is present in both star-forming and quiescent galaxies but is particularly dramatic in the latter case.  

Part of the observed trend in median size is due to the continual addition of new, larger galaxies to the massive, quiescent population.  Star-forming disks have larger sizes at later times \citep{Mo1998}.  As these galaxies quench, their appearance in the quiescent population can shift the median size without changing the number density of compact galaxies, a phenomenon known as ``progenitor bias" \citep{vanDokkum2001, Carollo2013}. Careful accounting suggests that progenitor bias alone is insufficient to explain the observed size growth, however \citep{Belli2015, Keating2015}, so that the growth of individual galaxies within the quiescent population is also required.  Recently, some authors have found that compact galaxies may not be as rare in the local universe as previously thought \citep{Damjanov2014, Damjanov2015, Saulder2015}, suggesting that they must continue to be produced to replace any which ``size out" of the compact selection.

Several authors have proposed mechanisms for increasing galaxies' sizes after quenching.  Dry mergers between galaxies of similar masses increase size as $R_e \propto M$ \citep{Hernquist1993, Boylan-Kolchin2006, Hopkins2009d}, while minor mergers may increase the sizes even more efficiently, $R_e \propto M^{>1}$ \citep{Naab2009, Hopkins2010}.  In the latter case, it has been suggested that the massive, compact galaxies at high redshift now lie at the centers of the most massive galaxies in the local universe, having accreted a substantial amount of stellar mass since their initial formation \citep{Hopkins2009, VanDokkum2014} or regrown a star-forming disk \citep{Graham2015}.  This idea is supported by the observational fact that massive galaxies' interior densities evolve much less strongly than their overall size \citep{Bezanson2009}.  The observed minor merger rate, however, may be too low to be entirely responsible for galaxies' growth \citep{Newman2012}. In the absence of mergers and accretion, galaxies may also ``puff up" due to adiabatic expansion in response to the expulsion of gas from the inner regions from stellar winds and black hole feedback as $R_e \propto M^{-1}$ \citep{Fan2008, Fan2010}.

These attempts to explain what happens to high-redshift compact galaxies are hindered by the long timescales of galactic evolution, which limit observations of any individual system to a single evolutionary state.  As a result, galaxy populations can only be connected at different epochs in a statistical sense.   Simulations of galaxy formation, however, offer the opportunity to directly ``observe" the evolution of individual systems over time.  Large-volume simulations, moreover, can provide a cosmological context for that evolution and produce populations of galaxies which can be compared against observations.

\begin{figure*}
  \centering
  \includegraphics[width=2.1\columnwidth]{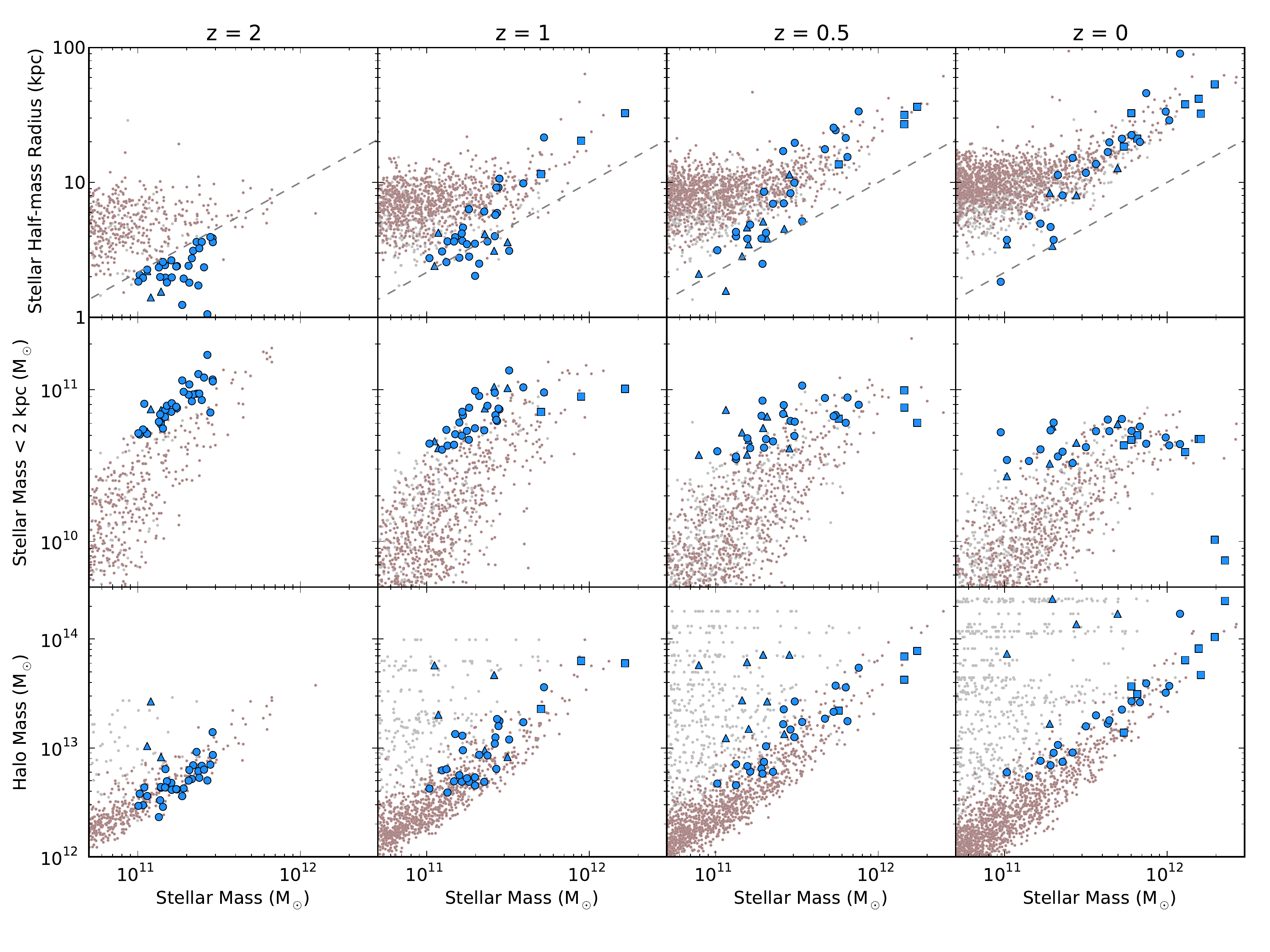}
  \caption{Evolution of the $z=2$ compact population to the present day.  Throughout, the compact galaxies' descendants are shown in blue symbols.  Triangles indicate satellite galaxies at the current redshift, and squares indicate that the compact galaxy is not in the main progenitor branch of its descendant's merger tree (i.e., that it merged with a more massive galaxy).  The remainder of the massive galaxy population is shown in dark pink and light grey points indicating central and satellite galaxies respectively.   {\it Top row:} Size-mass plane at $z$=2, 1, 0.5, and 0. The $\log(\Sigma_{1.5}) > 10.5$ compactness selection criteria is represented by a dashed grey line.  {\it Middle row:} Total stellar mass within 2 (physical) kpc of the galaxy center. {\it Bottom row:}  Stellar mass - halo mass.  For satellite galaxies, the halo mass shown is that of the FOF group to which it belongs, so that it lies to the left of its central galaxy.  The compact galaxies' descendants spread out by a factor $\ge$ 6 in both stellar mass and halo mass but remain some of the densest cores in the simulation.}
  \label{fig:generalevol}
\end{figure*}

In \citet{Wellons2015}, we identified a set of galaxies in the cosmological hydrodynamical simulation Illustris which are analogs to observed massive compact ellipticals, with stellar masses of $1-3 \times 10^{11}$ \Msun~and stellar half-mass radii of 2 kpc or smaller at $z=2$.  These compact galaxies formed naturally within the cosmological volume alongside non-compact galaxies at similar stellar masses, and we traced their progenitors back to higher redshift to discover how their formation paths differed.  We found that the galaxies all reached their compact state through some combination of assembling their stellar mass at very early times when the universe was extremely dense, and/or intense, centralized bursts of star formation usually triggered by gas-rich mergers.  Both in our work and in that of other simulators (see e.g. zoom-in simulations by \citet{Ceverino2015a} and \citet{Zolotov2015}), gas-rich environments are required to dissipate angular momentum and produce such compact objects.

We now turn our attention to the low-redshift outcomes of these massive, compact galaxies.  Our aim in this work is to show how a specific population of galaxies which were massive and compact at $z=2$ evolve forward to the present, describe their descendants, and characterize their evolutionary paths.  As before, we select a massive, compact population at $z=2$ in Illustris, this time tracing the galaxies forward to $z=0$.  In Section \ref{sec:illustris}, we describe the Illustris simulation suite and the merger trees used to connect galaxies at different redshifts.  In Section \ref{sec:general}, we select a population of 35 compact galaxies at $z=2$ and follow them forward to their $z=0$ descendants.  In Section \ref{sec:descendants}, we take a detailed look at four distinct types of descendants and show how common each type of evolutionary track is, and in Section \ref{sec:environment} we discuss how a galaxy's $z=2$ environment might influence its eventual outcome.  Finally, in Section \ref{sec:discuss} we address the likelihood of massive galaxies in the local universe hosting compact cores, explore the different mechanisms for size growth, and discuss the implications for the number density of compact galaxies at low redshift.  We summarize the results and conclude in Section \ref{sec:conclusion}.

\section{Illustris}
\label{sec:illustris}

Illustris \citep{Vogelsberger2014a, Vogelsberger2014, Genel2014, Nelson2015} is a cosmological hydrodynamical simulation with a periodic volume of (106.5 Mpc)$^3$ that is initialized at $z=127$ using cosmological intial conditions and evolved forward to the present day.  The simulation suite consists of 6 runs in total: 3 different levels of resolution, each run once with dark matter particles only and once with full baryonic physics, including gas cells, stellar and wind particles, and black hole particles.  The results presented in this paper are drawn from the highest-resolution full-physics run, Illustris-1, which has a baryonic mass resolution element of about $10^6$ M$_\odot$.  The gravitational forces exerted by DM and star particles are softened on a comoving scale of $\epsilon$ = 1.4 kpc.  For gas cells, the force softening is spatially adaptive with cell volume.  For all baryonic resolution elements (including both star particles and gas), $\epsilon$ is limited to a maximum physical size of 0.7 kpc at $z=1$.  

In addition to gravity, the full-physics runs of Illustris include a hydrodynamical treatment of gas using the moving-mesh code {\sc AREPO} \citep{Springel2010} and models for the baryonic physical processes that regulate the growth of stellar mass.  These include primordial \citep{Katz1996} and metal-line gas cooling \citep{Wiersma2009a} with self-shielding corrections \citep{Rahmati2013}, star formation and evolution, supernova feedback \citep{Springel2003}, chemical enrichment of the ISM \citep{Wiersma2009}, and supermassive black hole growth with quasar-mode \citep{DiMatteo2005, Springel2005} and radio-mode feedback \citep{Sijacki2007, Sijacki2015}.  These models are described in detail by \citet{Vogelsberger2013} and \citet{Torrey2014}, who show that they produce realistic galaxy populations and reproduce key observables across cosmological time.

 Of particular relevance to the results presented herein are the models for star formation, stellar feedback, and AGN feedback, which affect the sizes and quenching of galaxies.  Star-forming gas is modeled using a \citet{Springel2003} effective equation of state which forms stars stochastically once above a certain density threshold, according to a \citet{Chabrier2003} initial mass function.  As stars evolve via winds and supernovae, mass and metals are returned to the gas as wind particles launched in a bipolar outflow with velocities scaled to the local DM velocity dispersion.  Supermassive black holes accrete mass from the surrounding gas at a rate that scales with the gas density and black hole mass and limited by the sound speed of the gas.  At high accretion rates, black hole feedback operates in a quasar mode which thermally returns a portion of the accreted energy to the nearby gas, and at low accretion rates black hole feedback occurs mechanically by the injection of hot radio bubbles whenever the black hole has accreted a small additional fraction of its own mass.   For more details, we refer the reader to \citet{Vogelsberger2013}.  These physical models, in combination with those mentioned above, create galaxy populations which reproduce the stellar mass function, cosmic star formation rate, black hole mass scalings, and a mixture of galaxy morphologies and colors as observed in the real universe.  

The simulation does deviate from observations in a few ways that are relevant to this work -- in particular, the size-mass relation is too shallow at stellar masses $\lesssim 10^{10.5}$ \Msun~(i.e., galaxies below this mass have sizes which are too large), and the quenching of star formation appears to be incomplete at the high-mass end (i.e., the color bimodality is not as strong as observed, which is due to a low level of residual star formation post-quenching), see \citet{Nelson2015} for details.  The galaxies we focus on in this paper, however, are only lightly affected by these limitations as they lie in a mass regime where the sizes are in reasonable agreement with observations, and in most cases do not undergo much star formation at low $z$.  Additionally, the softened gravity employed in Illustris may result in lower central gas densities and lower central star formation rates, hence the stellar masses quoted herein may be lower, and the stellar half-mass radii larger, than they would be otherwise (see \citet{Sparre2015} and the convergence study in \citet{Wellons2015}).

The {\sc SUBFIND} algorithm \citep{Springel2001, Dolag2009} is employed to identify gravitationally-bound structures (i.e. DM halos and subhalos) within the simulation volume.  Their baryonic components (gas, stars, and black holes) comprise the galaxies that inhabit each subhalo.

Subhalos are tracked from snapshot to snapshot using the {\sc SUBLINK} merger trees developed by \citet{Rodriguez-Gomez2015}.  
A subhalo's potential descendants in the following snapshot are identified by shared ownership of individual particles (DM, stars, and gas).  The particles are ranked according to gravitational binding energy and weighted according to (rank)$^{-1}$.  The subhalo with the highest weighted sum over all shared particles is then assigned as the descendant.
A merger occurs when two or more subhalos share a common descendant.  
The main (or primary) progenitor of this descendant is chosen according to which subhalo has the ``most massive history" \citep{DeLucia2007}, which protects against the misallocation of particles that can occur when subhalos are close to merging.  We traverse this merger tree forward in time to find the low-redshift descendants of our compact galaxies.

\section{Evolution of the compact population}
\label{sec:general}

\subsection{Sample selection at $z=2$}
\label{ssec:selection}

We quantify ``compactness" using the observationally-motivated \citep[see e.g.][]{Barro2013} quantity
\begin{equation}
\Sigma_{1.5} = M_* / R^{1.5}
\end{equation}
where $M_*$ is a galaxy's stellar mass and $R$ its size.  This constitutes a diagonal cut across the mass-radius plane so that lower-mass galaxies must have smaller sizes in order to qualify as compact.  We use subhalo stellar masses and stellar half-mass radii to select a population of galaxies at $z=2$ with 
\begin{equation} 
\log \left( \frac{\Sigma_{1.5}}{{\rm M}_\odot / {\rm kpc}^{1.5}} \right) > 10.5
\end{equation}
and stellar masses of $1-3 \times 10^{11}$ M$_\odot$, consisting of 35 galaxies in total.  This selection differs from (and is more generous than) the strict 2 kpc size threshold imposed in \citet{Wellons2015}.  With this $\Sigma_{1.5}$ criterion, a galaxy of $10^{11}$ M$_\odot$ must still have a size of approximately 2 kpc or below, but we now allow larger sizes at higher masses so the total size of the compact population is more than doubled from the previous work.  (The original 14 galaxies are a subsample of the 35 galaxies we examine here.)  The selection criteria and resulting compact population are shown on the $z=2$ size-mass plane in the upper left-hand panel of Figure \ref{fig:generalevol}.   Note that this mass cut selects galaxies which already had stellar masses of $10^{11}$ \Msun~at $z=2$, so that the results herein pertain only to high-mass compact galaxies.  Two-thirds of the compact galaxies thus selected are quiescent (sSFR $< 3 \times 10^{-10}$/yr) at $z=2$, which is a broadly similar fraction as is found in observations \citep{Barro2014, VanDokkum2015}.

Note that our definition of $\Sigma_{1.5}$ is not directly comparable to that used by observers, since we measure mass and radius in different ways.  Stellar masses are computed as a sum of the masses of all star particles bound to the subhalo rather than through SED fitting, though these masses are generally the same to within a factor of two \citep{Torrey2014a}.  The radius used here is the stellar half-mass radius (a three-dimensional quantity), rather than a projected two-dimensional half-light or effective radius \citep[which will be smaller by about 25\% on average, see e.g.][]{Hernquist1990}.  In \citet{Wellons2015}, we measured effective radii from Sersic fits to mock observations for a small number of compact galaxies and found that the measured effective radii were 0.5-1 times the physical 3D radii, with an average difference of 0.74 as expected from projection.  The use of 3D radii therefore has the effect of making our values of $\Sigma_{1.5}$ smaller by $\sim0.2$ dex.  The threshold compactness value of 10.5 that we employ here is thus more restrictive than the cutoff of 10.3 using 2D sizes imposed by \citet{Barro2013} by roughly $0.4$ dex, selecting for the very most compact galaxies in the simulation volume.  (See Section \ref{ssec:generalize} for a discussion of how our results depend on the choice of threshold.)

\subsection{Sample evolution to $z=0$}

The descendants of these compact galaxies are traced forward in time using the merger trees described in Section \ref{sec:illustris}.  Figure \ref{fig:generalevol} shows the properties of these descendants at $z=2, 1, 0.5$ and 0.  Satellite galaxies at each redshift are marked with triangles, and a square denotes that the compact galaxy is not in the main progenitor branch of its descendant's merger tree at that redshift, i.e. that it merged with a more massive galaxy. 

Comparing the $z=0$ and $z=2$ panels, a couple of conclusions can immediately be drawn.  First, the compact galaxies' $z=0$ descendants are dispersed widely among the massive galaxy population by a factor of $\sim$20 in stellar mass despite having been clustered to within a factor of 3 at $z=2$.  Some galaxies maintain a similar mass all the way to the present day, while others grow by a factor of 10 or more.  This order of magnitude in the spread of galaxies' descendant masses implies that rank order cannot be preserved, a topic which we discuss in more detail in \citet{Torrey2015}.

Additionally, when comparing the numbers of galaxies that fall below the $\log(\Sigma_{1.5}) > 10.5$ line, it is apparent that there are many fewer compact galaxies by $z=0$ than there were at $z=2$.  As we will show in the following sections, processes that galaxies undergo at low redshift tend to increase their sizes as well as their masses.  As a result, we predict that the comoving number density of massive, compact galaxies should decrease in the local universe (see Section \ref{ssec:numberdensities}).

The half-mass radius, by its mathematical definition, can be increased by 
(i) physically inflating the profile itself (i.e. shifting stellar mass to larger radii and decreasing the density everywhere) 
(ii) expelling mass from the inner regions, or 
(iii) adding mass to the outer regions.  In the latter case, the half-mass radius will change while the core density remains the same, while the former two will also decrease the core density.  In the middle row of panels in Figure \ref{fig:generalevol}, we show the stellar mass within 2 physical kpc of the galaxy center (the core mass) at the same set of snapshots.  By definition, the compact galaxies sit on the high end of the distribution of core masses at $z=2$.  As time goes on, the typical core mass decreases for the compact galaxies as well as for the overall galaxy population. By $z=0$, the compact galaxies' descendants' cores are less dense than their progenitors' by about a factor of two on average, but they are still among the densest stellar cores in the simulation.  (The two obvious exceptions to this rule are massive galaxies which experienced strong interactions between $z=0.5-0$ that disrupted their cores.) The evolution in core density is less pronounced than the evolution in overall size, indicating that while the first two mechanisms must be at work, the third is driving much of the total size growth.

\begin{figure*}
  \centering
  \includegraphics[width=1.8\columnwidth]{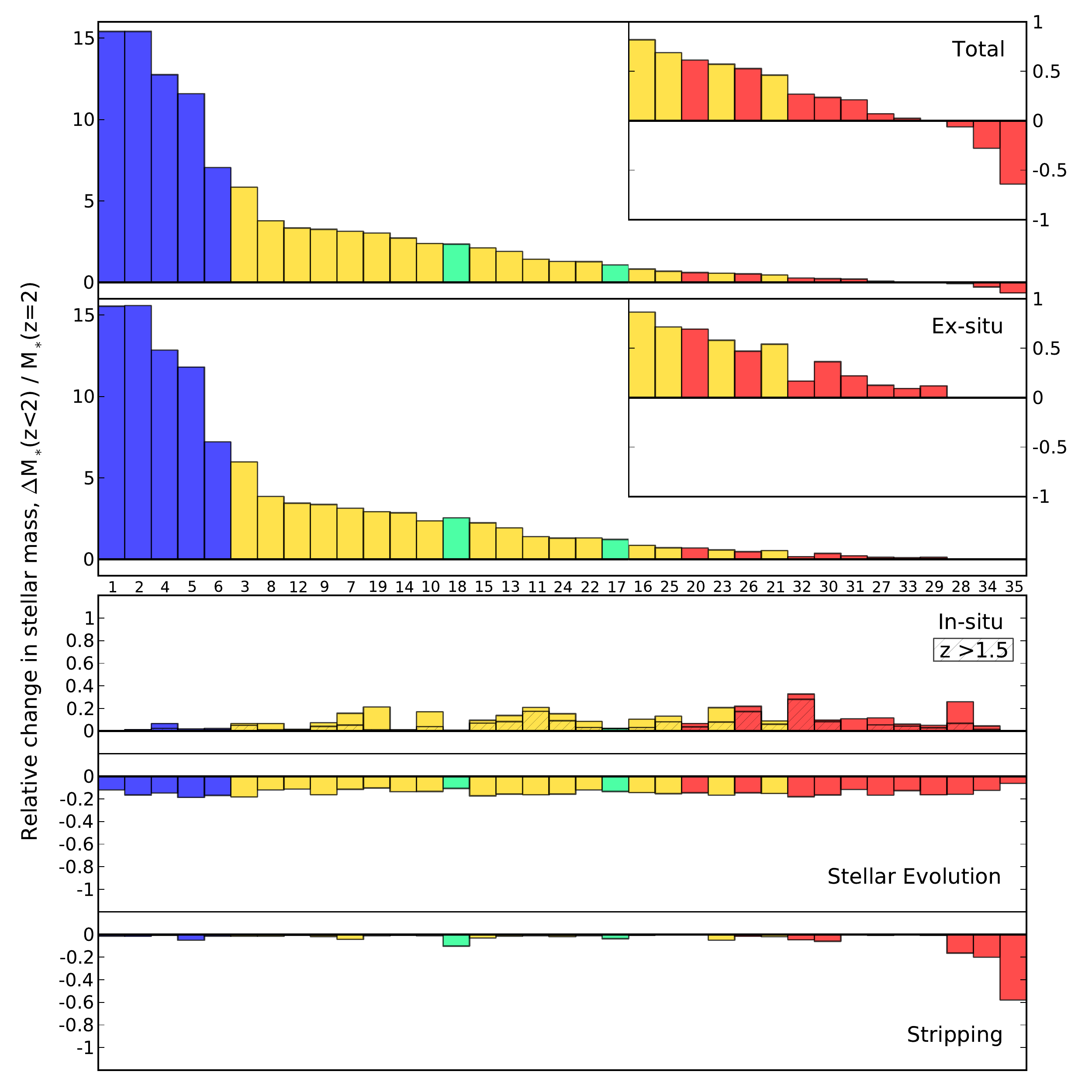}
  \caption{The amount by which each compact galaxy's stellar mass changes between $z=2$ and $z=0$ relative to its $z=2$ mass {\it (top panel)}, broken down into the contributions from ex-situ (i.e. accreted) stars {\it (second panel)}, in-situ star formation {\it (third panel)}, stellar mass loss {\it (fourth panel)}, and stripping {\it (bottom panel)}.  In the third panel, in-situ star formation which occurred between our selection redshift of $z=2$ and $z=1.5$ is indicated by hashed regions.  The galaxies are ordered by decreasing change in stellar mass, and their index in Figure \ref{fig:imagegrid} is noted in the middle of the figure.  Colors indicate the descendant class from Section \ref{ssec:categorize}.  Note that the top two panels are scaled differently from the bottom three, with insets on a comparable scale when the change in mass drops below 100\%.  In general the change in mass is dominated by ex-situ stars, but at the low-mass end where the galaxies change their mass by less than 50\%, other effects become significant.}
  \label{fig:deltaM}
\end{figure*}

The masses of the dark matter halos in which the galaxies are embedded are shown in the lower set of panels of Figure \ref{fig:generalevol}, with the stellar mass -- halo mass relation appearing across the diagonal in dark pink points.  The light grey points which lie above correspond to satellite galaxies.  At $z=2$, the compact galaxies are spread evenly across the scatter in the relation.  At lower redshifts, the compact galaxies which still remain in the original stellar mass bin become skewed toward higher halo masses, suggesting that while the galaxies themselves are quiescent, their dark matter halos do continue to accrete and grow in mass.  Similar to the stellar masses, the original factor of $\sim5$ spread in halo mass at $z=2$ has increased to a factor of 40 by $z=0$.

One of the few quantities that does {\it not} experience much dispersion from $z=2-0$ is the mass of the galaxies' supermassive black holes, $M_{\rm BH}$.  At $z=2$, the compact galaxies host black holes with $M_{\rm BH} = 0.6-3 \times 10^9$ \Msun, which is overmassive by about a factor of 2.5 in comparison to the median $M_{\rm BH}$ in the $M_* = 1-3 \times 10^{11}$ \Msun ~range in the simulation.  By $z=0$, the descendants' BH masses have approximately doubled to $1-6 \times 10^9$ \Msun, with a shallower than linear relation between the descendants' stellar and BH masses.  At all descendant stellar masses, these BHs are on the high edge of the $M_*$ - $M_{\rm BH}$ relation, and at lower stellar masses they become more extreme outliers.  At the lowest descendant stellar masses in our study of 35 galaxies, the descendants that still lie in the same range $M_* = 1-3 \times 10^{11}$ \Msun ~host BHs with masses of $1-2.6 \times 10^{9}$ \Msun, which is a factor of $\sim 8$ higher than the median $M_{\rm BH}$ in that range at $z=0$ in the simulation.  As discussed in \citet{Wellons2015}, in the models employed by Illustris both the black hole accretion and central star formation rate are functions of the density and temperature of the central gas, and so the black hole accretion rate tends to track the central star formation rate.  Thus, galaxies which are dominated by stars formed close to the center will also tend to have high black hole masses.  These results echo observations by \citet{Ferre-Mateu2015}, who find that ``\"uber-massive" SMBHs in the local universe are hosted by massive, compact relic galaxies with old stellar populations.

\begin{figure*}
  \centering
  \includegraphics[width=2.1\columnwidth]{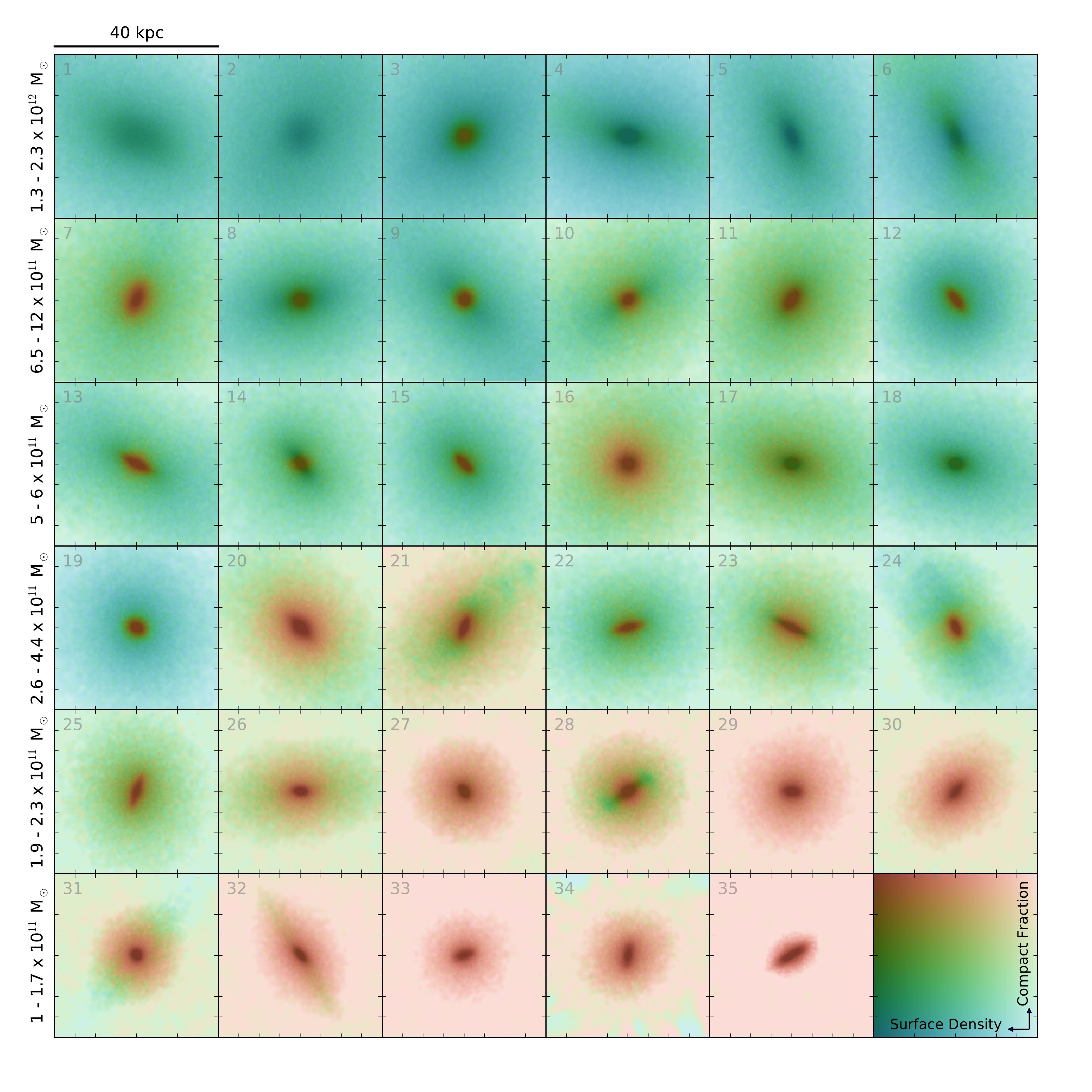}
  \caption{Images of the $z=0$ descendants of each of the $z=2$ compact galaxies, ordered by descendant mass.  A color's intensity is determined by stellar surface density, and its hue by what fraction of the stars along the line of sight belong to the compact progenitor at $z=2$ (legend in bottom right).  Thus the least massive descendants which have primarily preserved their compact identity appear red, while the most massive descendants which are dominated by other stars appear blue.  In many cases, there is a color gradient of red to blue from the inner to outer regions of the galaxy indicating that the compact galaxy is concentrated at the core of its more massive descendant.}
  \label{fig:imagegrid}
\end{figure*}

\subsection{Stellar composition of descendants}

We have seen in Figure \ref{fig:generalevol} that the compact galaxies can grow in stellar mass by varying amounts between $z=2$ and $z=0$, some gaining no mass at all while others grow by a factor of 10 or more.  In Figure \ref{fig:deltaM}, we show a detailed accounting of what drives this change for each galaxy.  In each panel, the galaxies are ordered by their total change in mass relative to their $z=2$ mass (the quantity shown in the top panel).  The panels below break down that total change into the contributions from {\it(i)} ex-situ stellar mass formed elsewhere and brought in by mergers or accretion since $z=2$, {\it(ii)} stars formed in situ since $z=2$, {\it(iii)} the decrease in stellar mass over time due to the death of massive stars, and {\it(iv)} the loss of stellar mass due to stripping.  (At the top of the figure, the numbers show the index of each galaxy when they are ordered by absolute descendant mass, corresponding to figures elsewhere in the paper.  Colors correspond to the type of descendant, which will be discussed in Section \ref{ssec:categorize}.)  The classification of star particles as in-situ or ex-situ is identical to that of Rodriguez-Gomez et al. (to be submitted): a particle is considered to have formed in situ if the galaxy in which it formed lies along the main branch of the descendant's merger tree.  Otherwise, it is considered ex-situ.  In cases where the compact progenitor itself does not lie along the main branch of its descendant at $z=0$, we identify stars formed in situ using the classification from the last snapshot when the compact progenitor {\it did} lie along the main branch.  A star particle is considered to have been stripped if it was present in the compact galaxy at $z=2$ but absent at $z=0$.

Note that the first and second panels for the total change in stellar mass and the mass of ex-situ stars are scaled to 15 while the remaining three panels are scaled to 1, and that they appear very similar except at the far right of the figure.  When the stellar mass changes by more than about 50\% (which is the case for 75\% of the compact galaxies), the change is completely dominated by the acquisition of ex-situ mass.  Insets in the top right of these panels zoom in on the low-mass end on the same scale as the bottom three panels.

Below, the middle panel shows the relative stellar mass growth due to in-situ star formation after $z=2$.  Several of the compact galaxies were ``caught in the act" at $z=2$ and were still star-forming at that time, though they quench (i.e., reach specific star formation rates below $0.1$ Gyr$^{-1}$ and continue to fall) by $z=1.5$.  This residual star formation between $z=2$ and $z=1.5$ is marked by hashing, and a few galaxies (e.g. 11, 26, and 32) continued to form up to 20\% additional stellar mass during this time.  In most cases the galaxies remain quiescent after quenching, with a couple noticeable cases of rejuvenation (e.g. 19 and 28) which also gain them up to 20\% more stellar mass. 

The final two panels show means of losing stellar mass rather than gaining it.  The fourth panel shows the change in stellar mass due to stellar evolution, i.e. massive stars' post-main-sequence evolution and death via supernovae.  This results in a decrease in stellar mass of about 15\%, which is fairly constant across the compact population.   Finally, the bottom panel shows how much stellar mass the galaxies have lost since $z=2$ through stripping (i.e., the total mass of all star particles which were present at $z=2$ but are missing at $z=0$).  For most of the compact galaxies, stripping is inconsequential, but for the three galaxies at the low-mass end it is actually significant enough that they have a net loss of stellar mass.  In the most extreme case, galaxy 35 dove through the middle of another halo and was dramatically stripped of 60\% of its stars, leaving only the most bound part of the core behind as it emerges from the other side at $z=0$.  Galaxies 28 and 34 are both satellites at $z=0$ and have been more gently stripped through interactions with their central galaxy.

In general, the galaxies' change in stellar mass between $z=2$ and $z=0$ can be largely attributed to ex-situ mass acquired through mergers and interactions, with other effects being sub-dominant. 

\section{Characterizing the descendants}
\label{sec:descendants}

\subsection{Classifying by compact fraction}
\label{ssec:categorize}

In the previous section, we have seen that the compact galaxies, originally tightly grouped at $z=2$, can accrue widely differing amounts of stellar mass by $z=0$.  What events in a galaxy's evolutionary history drive that assembly of stellar mass, and where do the stars from the compact progenitor lie within the resulting stellar distribution?

Figure \ref{fig:imagegrid} displays images of the $z=0$ descendant of each of the 35 compact galaxies, ordered by decreasing descendant mass.  The color in a given pixel is determined by two quantities, stellar surface density and compact fraction, as shown in the legend in the bottom right panel.  The color's intensity (how light or dark it is) indicates the stellar surface density (the total mass of all the star particles along the line of sight) so that denser regions are darker.  The hue indicates how many stars in that pixel belong to the compact progenitor (the ``compact fraction") so that regions dominated by stars from the original compact galaxy are red and regions dominated by other stars are blue.  (In detail, the compact fraction is calculated as the fraction of stars in that pixel which were either present in the compact progenitor at $z=2$ or formed in-situ before $z=1.5$.)  In some cases (e.g. 5/6 and 17/18), two compact galaxies have the same descendant.  In the top row, the galaxies 5 and 6 were both consumed by the same massive galaxy, and in the third row galaxies 17 and 18 merged together.  In such cases, the two panels will then have identical surface densities but differing compact fractions, since they refer to different progenitors.

Seeing the descendants in this way again highlights the variety of experiences they undergo.  Near the top where the most massive descendants reside, in many cases the compact progenitor galaxy cannot be readily identified as a single object, indicating that it was tidally shredded through an interaction with a more massive galaxy.  In several of these, a band of green highlights the plane on which the compact galaxy was disrupted.  In others, the core of the galaxy is orange or red while the outskirts are blue, indicating that the compact galaxy was able to maintain coherence throughout the merger process(es) and exists as the core of its descendant.  Moving further down in mass, the panels which are primarily red in color are compact galaxies which experienced little merger or star formation activity between $z=2$ and $z=0$.

\begin{figure*}
  \centering
  \includegraphics[width=2.1\columnwidth]{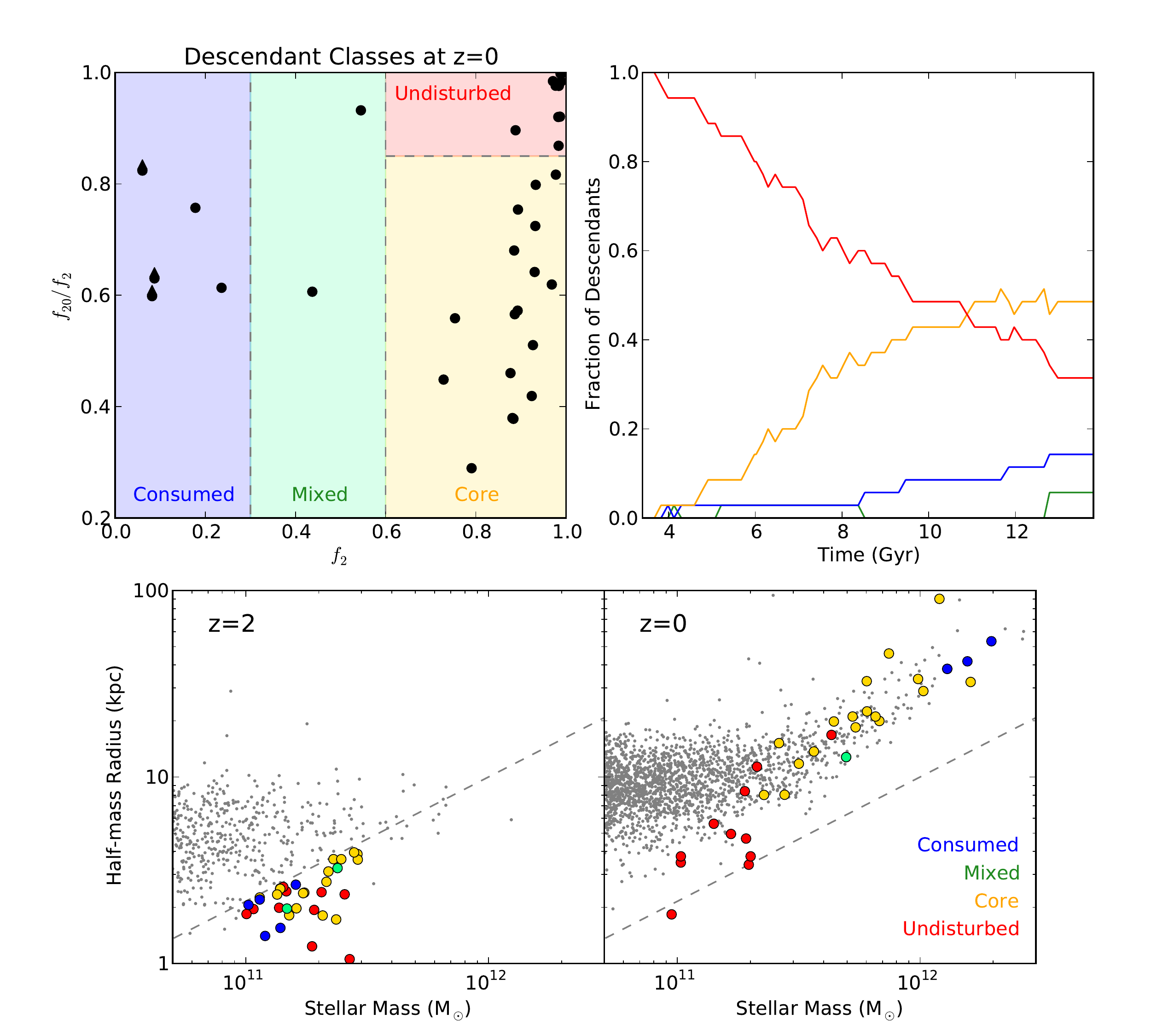}
  \caption{Classification scheme for the descendants of the compact galaxies, as described in Section \ref{ssec:categorize}, where $f_r$ is the fraction of stars within $r$ kpc that belong to the original $z=2$ compact progenitor.  {\it Top left:} Values of $f_2$ and $f_{20}/f_2$ for the final descendants at $z=0$, with grey dashed lines and shaded regions showing how they are classified.  An upward-pointing triangle indicates that $f_{20}/f_2 > 1$, so the point has been moved down by a factor of two (which does not affect the classification). {\it Top right:} The fraction of descendants in each class as a function of time, applying the same scheme to the descendants of the $z=2$ compact galaxy population at each redshift.  By definition, the galaxies are all initially ``undisturbed" at $z=2$, and fall into other classes as they acquire additional stellar mass. {\it Bottom:} The compact population at $z=2$ {\it (left)} and their $z=0$ descendants {\it (right)} on the size-mass plane, colored according to final descendant class.  In every panel, the color blue indicates galaxies which were ``consumed" by a larger galaxy, green indicates galaxies which were ``mixed" in major mergers, gold indicates galaxies which are the ``core" of a more massive descendant, and red indicates galaxies which survive ``undisturbed" to the present day.  (Note that there are fewer colored points at $z=0$ because some of the compact galaxies have merged together.)  The grey dashed line indicates our compactness criterion, $\log(\Sigma_{1.5}) > 10.5$.}
  \label{fig:categories}
\end{figure*}

A more quantitative view of the galaxies' structure can be found in the appendix Figure \ref{fig:profgridz0}, which shows the cumulative stellar mass profiles of the $z=0$ descendants broken down into its original, ex-situ, and in-situ components.  To classify the variety of descendants quantitatively, we define
\begin{equation}
f_r = \frac{M_*(<r, {\rm compact})}{M_*(<r, {\rm all})}
\end{equation}
the fraction of stellar mass interior to a radius $r$ belonging to the compact progenitor.  (This compact fraction is measured in the same way as in Figure \ref{fig:imagegrid}, but includes all stellar mass enclosed within $r$ rather than the stellar mass in a single pixel.)  We then measure $f_r$ at radii of 2 kpc and 20 kpc (chosen to be representative of the core and envelope respectively) and use these quantities to inform our classification.  A galaxy whose core is compact-dominated, for example, will have a high $f_2$.  If it also has a high $f_{20}$, it is essentially the same galaxy, while a low $f_{20}$ means that it has acquired an ex-situ envelope.  We thus define four general classes of descendant, as depicted at $z=0$ in the top-left panel of Figure \ref{fig:categories}:
\begin{itemize}
\item \textbf{ Consumed:} $f_2 < 0.3$, the compact galaxy's identity was erased by accretion onto a more massive descendant.
\item \textbf{ Mixed:} $0.3 < f_2 < 0.6$, the compact galaxy experienced major merger activity and is partially disrupted.
\item \textbf{ Cores:} $f_2 > 0.6$ and $f_{20}/f_2 < 0.85$, the compact galaxy exists as the core of its more massive descendant.
\item \textbf{ Undisturbed:} $f_2 > 0.6$ and $f_{20}/f_2 > 0.85$, the compact galaxy has entirely preserved its identity.
\end{itemize}

\begin{figure*}
  \centering
  \includegraphics[width=1.8\columnwidth]{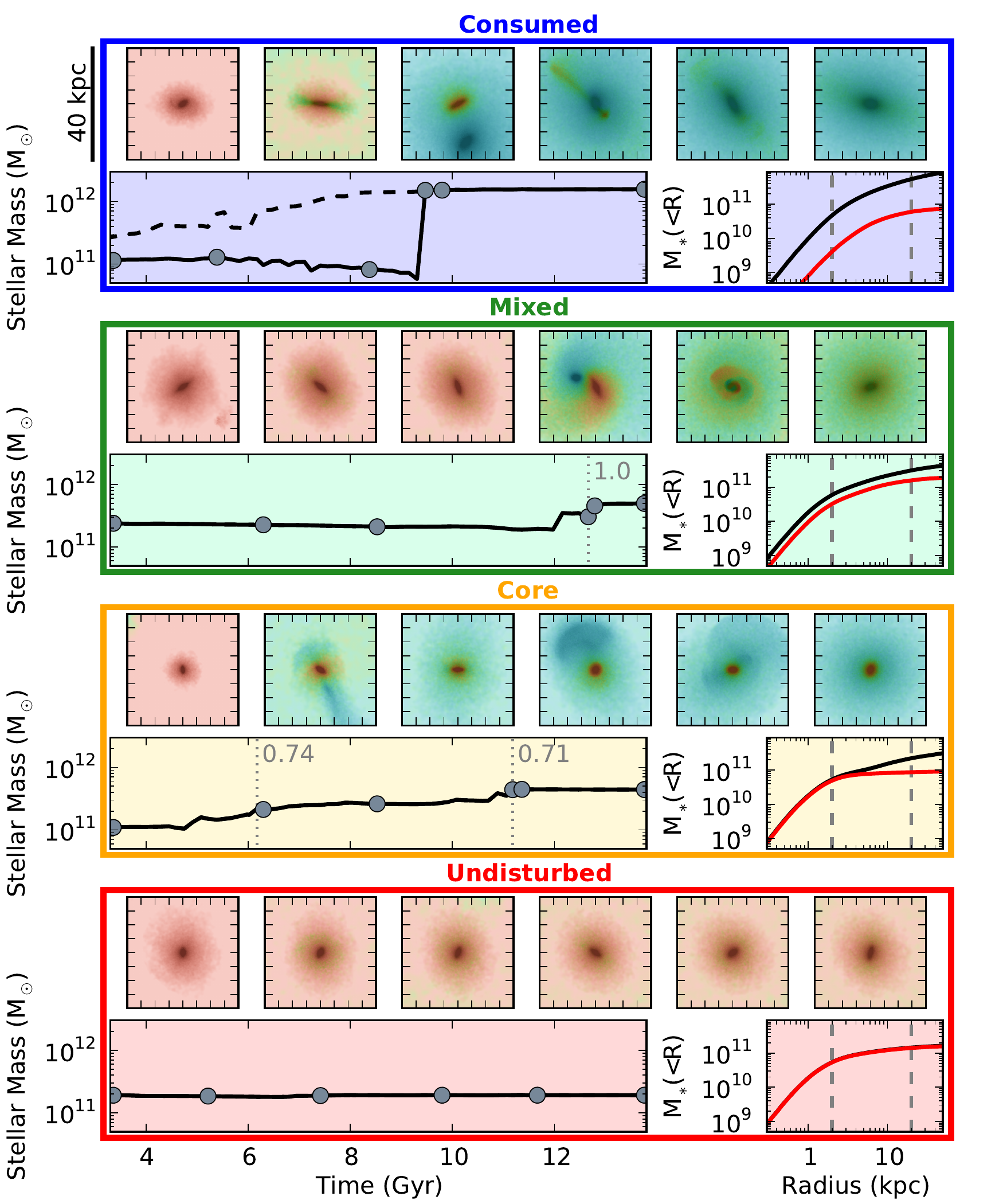}
  \caption{Evolution of an individual galaxy from each of the four descendant types.  (From top to bottom, these are galaxies 4, 17, 19, and 29.) For each of these examples, the panel on the left shows the evolution of stellar mass with time in a solid black line, with dashed lines drawn for merger partners where applicable.  Large grey points mark the times corresponding to the six images directly above, which have the same color scheme as in Figure \ref{fig:imagegrid}.  The rightmost panels show the cumulative stellar mass profiles for the final $z=0$ descendants.  Black lines include all stellar mass, while the red lines only include star particles that were present in the compact progenitor at $z=2$.  The ratio of these two lines is the $f_r$ discussed in Section \ref{ssec:categorize} used to classify the descendants, and grey dashed lines show where $f_2$ and $f_{20}$ were measured.  Each of these examples is demonstrative of the kind of events (or lack thereof) that characterize each descendant type.}
  \label{fig:indivexamples}
\end{figure*}

The specific choice of radii as well as the dividing lines between classes are somewhat arbitrary, particularly between the core and undisturbed classes.  However, we find that the classification is not very sensitive to our specific adopted radii as long as one radius is below 3 kpc and the other is at least 8 kpc.  Moreover, upon visual inspection of the mass profiles the appropriate class is obvious in most cases, so the boundaries can be chosen accordingly.  (See Figure \ref{fig:indivexamples} for archetypal examples in each class and Figure \ref{fig:profgridz0} for the full set of profiles.)  At $z=0$, of the original 35 compact galaxies, 5 have been ``consumed," 2 are ``mixed," about half are ``cores," and 11 are ``undisturbed."  The top-right panel of Figure \ref{fig:categories} shows how the classes developed over time.  All the galaxies are initially undisturbed (by definition), and over time the fraction of compact galaxies which develop outer stellar envelopes increases, leveling off at about 50\%.  Compact galaxies undergo mixing or consumption at a lower rate.  Note that galaxies may transition between classes more than once -- one galaxy, for example, is ``mixed" in a major merger and then later ``consumed."  We have found that using radii which scale with $z=2$ mass as $M_*^{1/1.5}$ produces the same results, so that within this small initial mass range the use of constant radii does not impose a mass-dependent bias.

The bottom panels in Figure \ref{fig:categories} show again the compact galaxies at $z=2$ and their descendants at $z=0$ on the mass-radius plane, this time colored according to the class to which the descendant belongs.  At the low-mass end at $z=0$, the descendants all belong to the ``undisturbed" class in red, since they have gained very little stellar mass.  At the high-mass end, those descendants which have consumed a compact galaxy are shown in blue.  Between those extremes lie compact galaxies which are cores (in orange) and which were thoroughly mixed by merging (in green).  (Note that there is not always a one-to-one correspondence between $z=2$ and 0 because some of the compact galaxies have the same descendant.)  Examples showing the evolution of an individual galaxy from each of these classes are shown in Figure \ref{fig:indivexamples} and discussed in the following subsection.

\subsection{Representative evolutionary examples}
\label{ssec:individs}

Each of the four sets of panels in Figure \ref{fig:indivexamples} is composed of three items demonstrating how a single compact galaxy from $z=2$ developed the characteristics that determined its descendant group at $z=0$.  The bottom left panel in each frame shows the evolution in stellar mass, with major mergers marked with a vertical dotted grey line and the stellar mass ratio listed at the top of the panel.  Large grey points correspond to the six images above, which are colored using the same scheme as in Figure \ref{fig:imagegrid} where a color's intensity indicates stellar surface density and hue indicates the fraction of stars belonging to the compact progenitor.  The image farthest to the left depicts the compact progenitor at $z=2$, and the image farthest to the right depicts its final $z=0$ descendant.  The cumulative stellar mass profile of this descendant is plotted in the panel in the lower right of each frame with a black line.  The red line, for contrast, shows the cumulative stellar mass profile of only those stars belonging to the compact progenitor.  The ratio of the red to black line is the quantity $f_r$ described in Section \ref{ssec:categorize}.  Grey dashed lines mark the locations of $f_2$ and $f_{20}$ which were used to classify the descendants.  Each of these examples demonstrates the kind of event (or lack thereof) that determines the final descendant type, and is representative of that class.

\begin{figure*}
  \centering
  \includegraphics[width=2.1\columnwidth]{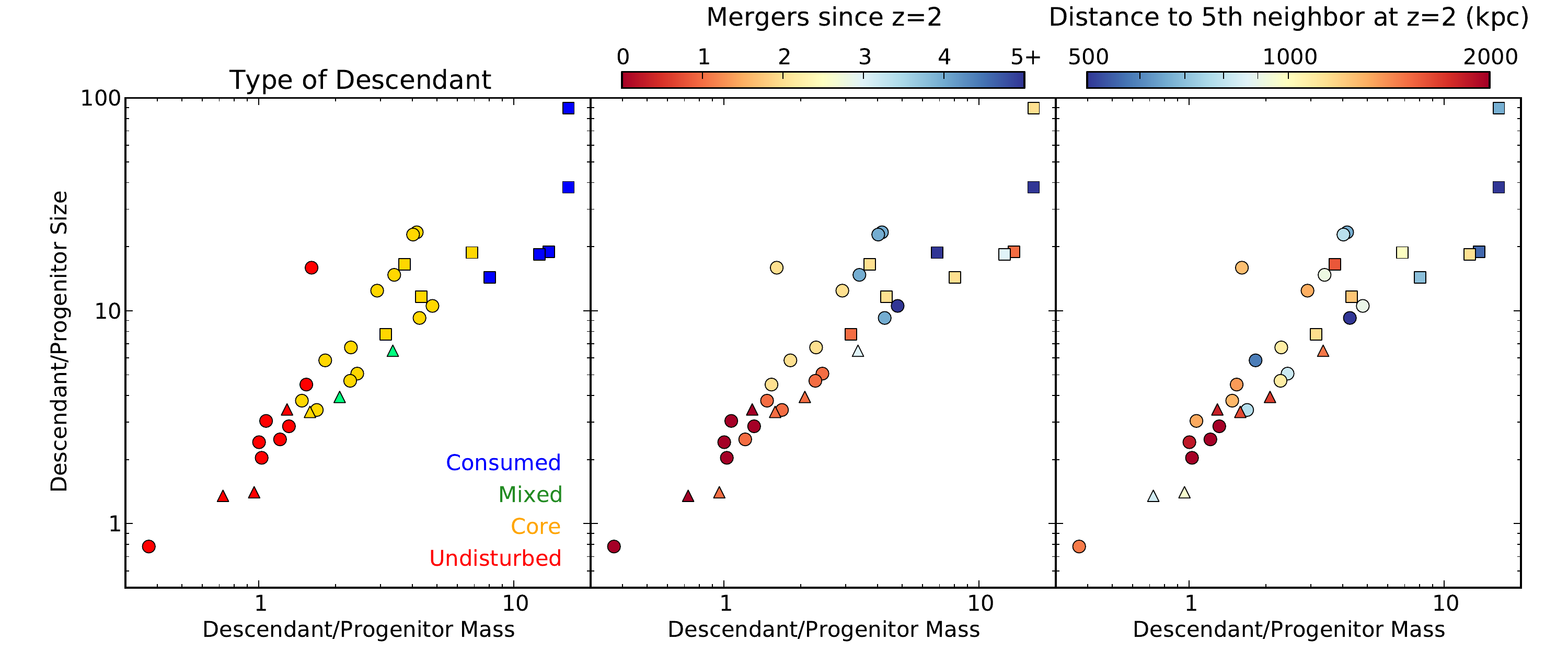}
  \caption{Growth in size and mass among the compact galaxy population from $z=2-0$.  In all three panels, the x-axis shows the mass growth (ratio of a galaxy's stellar mass at $z=0$ and $z=2$) and the y-axis shows the size growth (ratio of the stellar half-mass radii).  As in Figure \ref{fig:generalevol}, triangles indicate that a galaxy is a satellite at $z=0$ and squares indicate that the compact galaxy does not lie along the main progenitor branch of its descendant.  {\it Left panel:} color refers to the type of descendant, as defined in Section \ref{ssec:categorize}.  {\it Middle panel:}  color indicates the number of mergers since $z=2$ with stellar mass ratios of $<$10:1.  {\it Left panel:} color indicates $z=2$ environment, i.e. the distance to the 5th nearest neighbor with an r-band magnitude of at least -19.5 (about $10^9$ \Msun).  The most isolated galaxies (red on the rightmost panel) tend to experience fewer mergers (red on the middle panel) and thus experience little mass growth and are essentially undisturbed.  Two notable exceptions are satellite galaxies, which fell into another DM halo in the intervening time period but have not merged with the central galaxy.}
  \label{fig:environment}
\end{figure*}

\subsubsection{Consumed}
\label{sssec:consumed}

In the first (blue) frame in Figure \ref{fig:indivexamples} which shows the evolution of galaxy 4, the stellar mass shows a sudden jump around 9 Gyr which marks the point in time when the compact galaxy merges into a much more massive galaxy.  As it is falling in, its outer stellar envelope is slowly stripped away before the core finally falls into the central object. The images above show that the compact galaxy is completely torn apart and its identity erased, though it is still faintly visible at $z=0$ as a band of green tracing the orbital plane along which it was disrupted.  Five of the 35 compact galaxies (14\%) fall into this class at $z=0$.  Their $z=2$ progenitors tend to fall into the lower-mass end of the selection range, which is not surprising since the less massive a galaxy is, the more likely that a merger partner will be capable of disrupting it.  Interestingly (and perhaps unexpectedly), the regrowth of a star-forming disk is visible in the second image of the sequence.  This is not an uncommon event among the objects in our sample, with disk regrowth occuring at some point for about a quarter of them.  These disks are very minor in terms of their stellar mass when compared to the mass of the core, but would undoubtedly make up a more significant portion of the light during their star-forming period.

\subsubsection{Mixed}
\label{sssec:mixed}

The second (green) frame shows a galaxy (number 17) which was primarily undisturbed until about 12 Gyr.  At that time, it experienced a merger with another galaxy of similar mass.  As the two galaxies spiraled in together, their stars became thoroughly mixed.  As is visible in the profiles on the right, stellar material from the compact progenitor is distributed uniformly throughout the final descendant and makes up a large fraction of its total mass.  The two galaxies are able to mix together so efficiently because they have not only similar masses, but similar binding energies -- in fact, the galaxy which merges is also in the compact sample (number 18)!  As mergers between galaxies of nearly the same mass and binding energy are rare, galaxies 17 and 18 are the only two of the sample of 35 (6\%) which fall into this class of descendant.

\subsubsection{Cores}
\label{sssec:cores}

An example of the third and most common type of descendant is shown in the third (orange) frame.  This galaxy (number 19) experienced two major mergers between $z=2$ and 0, along with several more minor mergers.  At 6 Gyr, it disrupts another galaxy of 75\% its mass and gains a thin stellar envelope.  Later, around 11 Gyr, it does the same to another loosely bound massive galaxy at a similar mass ratio, ultimately residing at the core of a final descendant many times its original mass.  

These 3:4 merger events are not so different in terms of stellar mass ratio from the 1:1 merger in the previous (mixed) case.  The crucial difference between the two cases is not necessarily mass, but gravitational binding energy.  In dissipationless mergers, it has been shown that particles tend to preserve their rank order in binding energy, with the most tightly bound particles of the merger progenitors remaining the most tightly bound particles of the descendant \citep{Barnes1988, Hopkins2009d}.  The compact galaxies, with their large masses and small sizes, are very tightly bound.  Thus, they are more likely to survive a merger intact even in cases of near unity mass ratios, and it requires a galaxy of similar or greater mass to unbind these objects (as in the previous two examples).  17 of the original 35 compact galaxies (49\%) survive as the cores of more massive descendants in this manner.

\subsubsection{Undisturbed}
\label{sssec:undisturbed}

The evolution of the final object (galaxy 29, red frame) is remarkable in its uneventfulness.  This galaxy maintains the same stellar mass throughout its evolution, gaining stars neither through accretion, mergers, nor in-situ star formation.  Eleven objects from our sample of 35 (31\%) are similarly able to avoid further action and thus completely maintain their same identity from $z=2$ (though not all are quite so pristine).  Though these galaxies do not grow in stellar mass, they do grow in size, a topic which is further discussed in Section \ref{ssec:sizeevol}.

\section{Environment}
\label{sec:environment}

What determines which evolutionary path a compact galaxy will take?  Clearly, mergers are a very important component of their growth.  The compact galaxies are primarily quiescent at $z=2$ and retain low specific star formation rates throughout their lifetimes, so most of their mass growth is due to the acquisition of ex-situ stars through mergers and interactions.  Figure \ref{fig:environment} shows the relative growth in both mass ($x$-axis) and size ($y$-axis) for the compact sample between $z=2$ and 0.  In the first panel, galaxies are colored according to their descendant type, as in Figure \ref{fig:categories}. The bottom left region of this panel is dominated by the undisturbed (red) compact galaxies which have experienced little size and mass growth, while the rightmost galaxies are those which were consumed (blue) by their very massive descendants.  (The outlying red symbol which has grown significantly in size but not much in mass is galaxy 20, which has some ex-situ mass which was deposited at radii of tens of kpc or greater.  The steepness of the density profile at intermediate radii then forces the half-mass radius to expand substantially to enclose enough mass to compensate.)

In the middle panel, the symbol color indicates the number of mergers with stellar mass ratios of at least 1:10 that the galaxies have experienced since $z=2$.  The gradient in color from red to blue with increasing mass growth shows that, as expected, the more merger activity a galaxy experiences, the more stellar mass it gains and the more likely it is to fall into the ``core" or ``consumed" groups.  

\begin{figure}
  \centering
  \includegraphics[width=0.8\columnwidth]{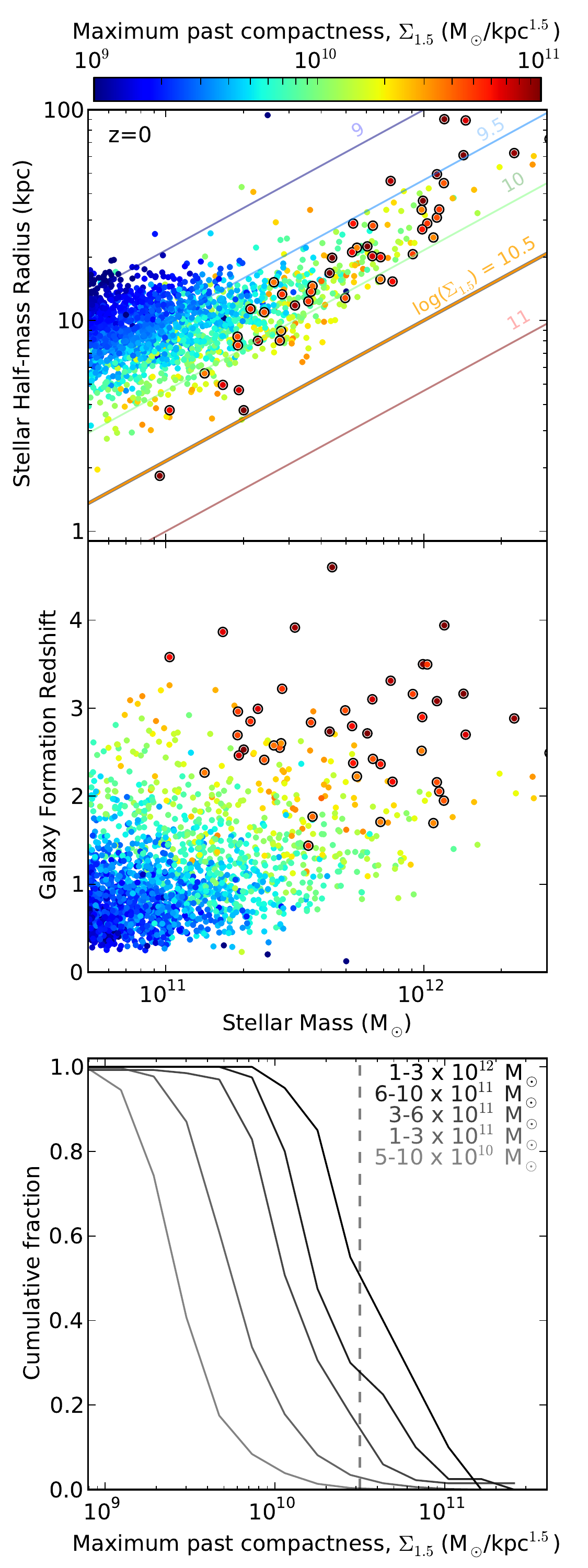}
  \caption{{\it Top panel:}  Mass-size plane at $z=0$ for Illustris galaxies.  Color indicates the maximum compactness ($\Sigma_{1.5}$) of all the galaxy's progenitors along the main branch of its merger tree above $10^{10}$ M$_\odot$.  Solid colored lines mark a constant value of $\Sigma_{1.5}$.  Galaxies which have satisfied $\log(\Sigma_{\rm max}) > 10.5$ (the compactness criterion employed in this paper) {\it at any redshift} are marked with black circles.  (Over half of these are the same 35 which were compact at precisely $z=2$.)  {\it Middle panel:}  Same as above, but with the y-axis indicating the galaxies' formation redshifts, defined as the median formation redshift over all in-situ stars.  {\it Lower panel:} Cumulative distribution of $\Sigma_{\rm max}$ for all galaxies, binned by stellar mass and normalized against the number of galaxies in each bin.  Thus, given a value of $\Sigma_{1.5}$ (e.g. $10^{10.5}$, marked with a vertical dashed line), one can determine what fraction of galaxies at a given stellar mass have ever achieved that level of compactness.  There is a clear relationship between stellar mass and past} compactness -- more massive galaxies have more compact progenitors -- but the trend has significant dispersion which is partially due to galaxies' differing formation times.
  \label{fig:maxpastcompact}
\end{figure}

In turn, a compact galaxy's merger rate is a function of its environment.  The rightmost panel of Fig. \ref{fig:environment} demonstrates this correlation by coloring the same points according to the density of galaxies in the compact progenitors' $z=2$ environments, quantified as the distance to the 5th nearest neighbor with an r-band magnitude of at least -19.5 (corresponding to a stellar mass of about $10^9$ \Msun) as in \citet{Vogelsberger2014a}.  The same trend appears in this panel as in the middle one, indicating that a compact galaxy's environment has some power to predict how it will evolve.  For this subsample of galaxies, it appears that in denser, more crowded environments, the galaxies are more likely to merge with their neighbors of comparable mass.  The more isolated (red) a compact galaxy is, the fewer mergers it will undergo, and the less mass growth it will experience.  The descendants of the galaxies which were in isolation at $z=2$ and experienced little merger activity in the interim are also preferentially isolated at $z=0$.  The connection between galaxy merger rate and environment, which was investigated here only for this particular set of galaxies which were massive, compact, and primarily quiescent at $z=2$, will be explored in more detail for all Illustris galaxies by Rodriguez-Gomez et al. (in prep).  

Two interesting exceptions to the merger rate/environment correlation are the white triangles in the lower left corner, which were in moderately dense environments at $z=2$ but experienced very little mass growth.  At $z=0$, these two are satellite galaxies which have fallen into a nearby massive dark matter halo in the intervening 10 Gyr but have not (yet) merged with the central galaxy.  A similar DM-only result is reported by \citet{Stringer2015}, who find that the most compact DM structures in the simulation are substructures embedded in larger halos.  Based on this evidence, we predict that massive compact relics in the local universe should preferentially live in regions where they are protected from mergers, either in isolation or as satellites in a larger group.  The latter option also has some observational support from \citet{Valentinuzzi2010}, who have found that nearby massive compact galaxies tend to live in cluster environments.  

\section{Discussion} 
\label{sec:discuss}

\subsection{Compact galaxies as cores}
\label{ssec:cores}

Given that about half of the compact galaxies from $z=2$ exist at $z=0$ as the cores of very massive galaxies, one can ask the reverse question: how many massive galaxies have cores which were once compact galaxies?  We address this question by tracing all the $z=0$ massive galaxies through the merger tree and calculating $\Sigma_{1.5}$ at each point on its main progenitor branch.  We then define the ``maximum past compactness" $\Sigma_{\rm max}$ to be the maximum value\footnote{Occasionally, {\sc SUBFIND} or the merger tree will misassign ownership of star particles or DM halos for a single snapshot, throwing off the calculated mass or size.  To minimize the effect of these glitches, we actually use the second-highest $\Sigma_{1.5}$.} of $\Sigma_{1.5}$ when the progenitor had a stellar mass of at least $10^{10}$ M$_\odot$.  In contrast to the analysis thus far, this approach permits a galaxy to be compact at any redshift, rather than specifically tying them to $z=2$.  This allows, for example, a galaxy which forms very early at $z=3-4$ in a compact state but grows out of the compact population by $z=2$, or a galaxy which forms via a compact starburst at $z=1.5$, both of which would be missed by the preceding analysis.

The colors of the points on the mass-size plane in the upper panel of Figure \ref{fig:maxpastcompact} indicate $\Sigma_{\rm max}$ for all of the massive galaxies at $z=0$.  Lines of constant $\Sigma_{1.5}$ are drawn in the same color scheme.  Any galaxy whose primary progenitor met the $\log(\Sigma_{1.5}) > 10.5$ compactness criterion at any point in time is outlined in black.  About half of these were compact at exactly $z=2$ and overlap with the sample discussed elsewhere in this paper, while the other half experienced their compact phases at other redshifts.  From this scatter plot, there is a clear trend between mass and maximum past compactness: more massive galaxies tend to have more compact progenitors.  (See also \citet{Genel2014} for a discussion of the relationship between mass and compactness in Illustris.)  Thus, many of the most massive galaxies were indeed once compact galaxies according to our definition -- but it is interesting to note that this is not universally the case, since there is also considerable scatter in the distribution of $\Sigma_{\rm max}$ at this mass (and, in fact, at all masses).

One cause of the spread in $\Sigma_{\rm max}$ at a given mass can be seen in the second panel, which shows the galaxies' formation times.  Here, we have defined ``formation time" as the median formation redshift of all star particles in the galaxy which formed {\it in-situ}.  (This excludes stars acquired via mergers.)  As before, color indicates $\Sigma_{\rm max}$.  In addition to the trend with stellar mass, a trend between formation time and past compactness is clearly visible -- this is a direct reflection of the fact that galaxies form with smaller sizes at earlier times.  Thus, the distribution of formation times for galaxies of a given mass also manifests as a distribution of past compactness.

The relationship between mass and $\Sigma_{\rm max}$ is further quantified in the lower panel of Figure \ref{fig:maxpastcompact}.  Each line shows the cumulative fractional distribution of $\Sigma_{\rm max}$ for the galaxies in a given stellar mass bin.  As was suggested by the upper panel, moving to lower masses shifts the distribution to lower past compactness values.  For any $\Sigma_{1.5}$, we can now see what fraction of galaxies at that mass have ever met that threshold.  The orange line denotes the $\log(\Sigma_{1.5}) > 10.5$ criterion used elsewhere in this paper.  Approximately 55\% of galaxies with $M_* > 10^{12}$ M$_\odot$ have met this definition of ``compact," along with 30\% of 6-10 $\times 10^{11}$ M$_\odot$ galaxies, 15\% of 3-6 $\times 10^{11}$ M$_\odot$ galaxies, etc.

Although a substantial fraction (55\%) of very massive galaxies are former compact galaxies, it is important to note that few of these were compact at exactly $z=2$.  Rather, their compact phases are spread over $z=1-3$, so that only about $15\%$ of galaxies above $10^{12}$ \Msun~at $z=0$ have compact progenitors at $z=2$.  Similarly, of the 35 $z=2$ compact galaxies, only 3 (9\%) have direct descendants in that mass range at $z=0$.  These fractions demonstrate once again that while there is indeed a strong trend between final stellar mass and past compactness, there is also a large dispersion in galaxies' evolutionary tracks that cannot be neglected.

\begin{figure*}
  \centering
  \includegraphics[width=2.1\columnwidth]{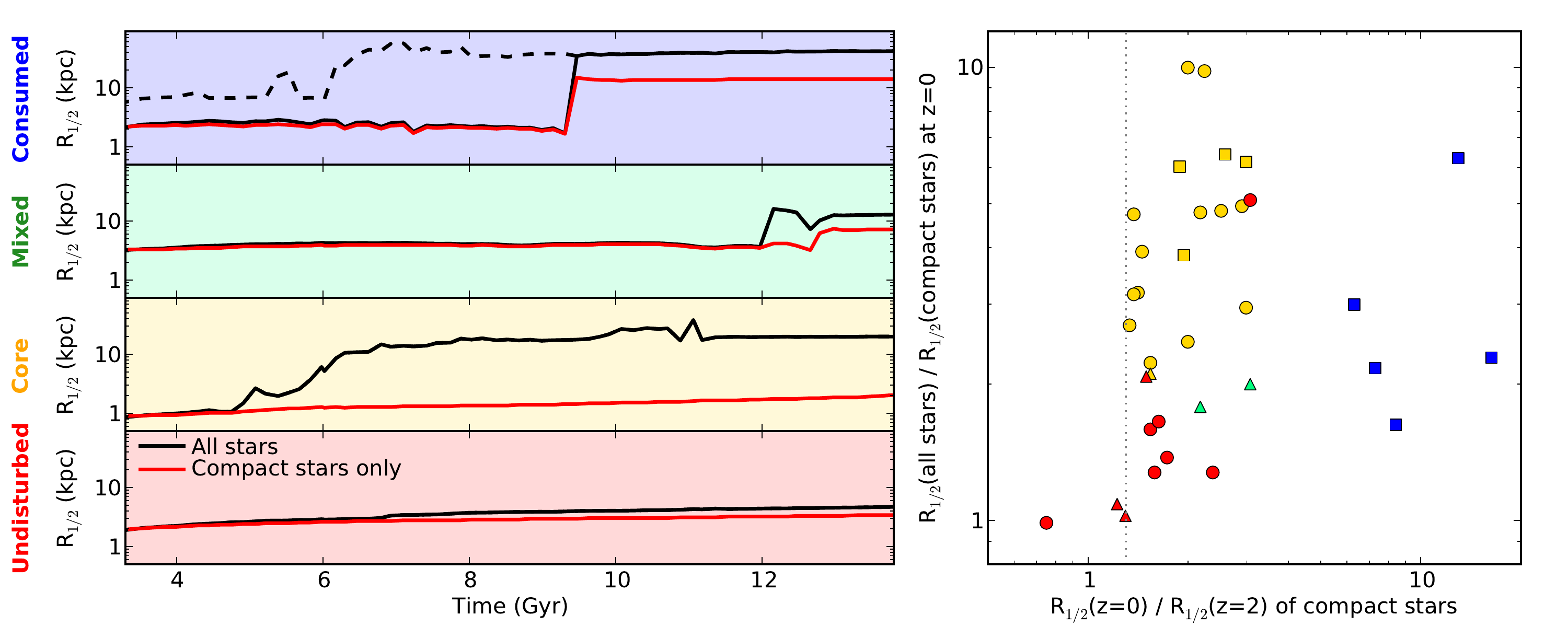}
  \caption{Compact galaxies' size evolution from $z=2-0$.  {\it Left panels:} Size evolution of the same four galaxies shown in Fig. \ref{fig:indivexamples}, one from each descendant group.  Black lines track the half-mass radius of the descendants' entire stellar distribution, while the red lines track the half-mass radius of only those stars which were part of the $z=2$ compact progenitor.  In the ``consumed" and ``mixed" cases, disruptive mergers increase the compact half-mass radius along with the overall half-mass radius.  In the ``core" case the compact galaxy remains essentially the same size while the overall half-mass radius increases due to a buildup of additional stellar mass at larger radii, and in the ``undisturbed" case the compact galaxy changes very little, having accreted or formed a very small amount of stellar mass since $z=2$.  {\it Right panel:}  For each of the descendants of the 35 compact galaxies, {\it x-axis:} Growth in half-mass radius of the compact progenitor's stars, i.e. the ratio between the value of the red line at $z=0$ and $z=2$.  The vertical dashed line shows the estimated 30\% growth from mass loss and numerical effects (see Appendix).  {\it y-axis:} Half-mass radius of all stars relative to the half-mass radius of compact stars, i.e. the ratio between the black and red lines at $z=0$.  Higher values in the $x$-direction indicate disruption of the compact core, while higher values in the $y$-direction indicate the presence of ex-situ stars at large radii.  Symbol shapes and colors are the same as in Figs. \ref{fig:categories} and \ref{fig:environment}.}
  \label{fig:sizeexamples}
\end{figure*}

\subsection{Mechanisms for size growth}
\label{ssec:sizeevol}

Nearly all of the compact galaxies have grown in size as well as mass since $z=2$, with very few still remaining in the compact regime at $z=0$.  Potential origins of this growth include physical effects such as major mergers \citep{Hernquist1993, Hopkins2009d}, minor mergers and accretion \citep{Naab2009, Hopkins2010}, star formation \citep{Graham2015}, and adiabatic expansion due to the expulsion of gas from the galaxy through stellar winds and BH feedback \citep{Fan2008, Fan2010}, as well as numerical effects stemming from the non-constant physical size of the gravitational softening length in Illustris.

To examine the importance of each of these mechanisms for the size growth of the different descendant types, let us return to the same four example galaxies we discussed in Section \ref{ssec:individs}.  The left-hand panels of Figure \ref{fig:sizeexamples} show the evolution of their stellar half-mass radii with time.  In addition to the black line for total stellar half-mass radius (which includes all stars bound to the object), we also show a red line for the the ``compact half-mass radius" -- that is, the half-mass radius including only those stars which belonged to the compact progenitor at $z=2$.  This allows us to disentangle the growth of the compact core from the growth of the galaxy as a whole.

In the first (consumed) case, both the overall and compact half-mass radii jump up when the compact galaxy falls into the more massive central galaxy.  This indicates, as we have seen, that the compact galaxy's stellar mass has been dispersed throughout the descendant.  In a similar (though less extreme) way, the second (mixed) galaxy's compact half-mass radius increases abruptly when it undergoes a disruptive major merger\footnote{The increase in total half-mass radius slightly ahead of the increase in compact half-mass radius marks the moment when {\sc SUBFIND} assigned the merger partner's outer stellar layers to the compact galaxy, prior to the actual merger event.}.  In these cases, the size growth is primarily driven by merging.

Conversely, in the third (core) case the compact half-mass radius hardly changes despite plenty of mergers and stellar mass growth, while the total half-mass radius greatly increases.  The very small change in compact half-mass radius signifies that the core has remained intact, and that the vast majority of the overall size growth is driven by the acquisition of outer layers of stellar material from less-bound galaxies. 

Finally, the galaxy which is undisturbed grows only slightly in size from 2 to 3 kpc over this 10 Gyr period.  In this case, the stellar mass has changed very little as a small amount of new stars ($\sim$10\%, both in-situ and ex-situ) offset the mass loss from preexisting stars.  Here, size growth mechanisms such as adiabatic expansion from mass loss and numerical effects start to become significant.  In treating the gravitational forces between baryons, Illustris uses a gravitational softening length $\epsilon_{\rm b}$ which is a fixed comoving size at redshifts $z \ge 1$.  The increase of the physical scale on which the potential is softened will decrease the depth of the gravitational well in dense regions, artificially inflating the stellar mass profiles as the star particles become less gravitationally bound.  It is unlikely that the increasing $\epsilon_{\rm b}$ is the sole culprit, however, since it is capped at a fixed physical size at $z=1$ but the size growth does not completely halt at this time.  In the Appendix, we estimate how these compact stellar systems will expand adiabatically in response to the changing softening length, stellar mass loss, and gas expulsion.  We find that numerical ``relaxation" of the galaxies' $z=2$ profiles to the $z=0$ softening length should only change the half-mass radius by about $15\%$.  Adiabatic expansion as a response to stellar mass loss and gas expulsion should produce another $\sim15$\% growth, and the remaining $\sim20$\% growth is owed to the small amount of in-situ stars formed and ex-situ stars deposited at larger radii.

Returning to the compact population as a whole, the right-hand panel of Figure \ref{fig:sizeexamples} shows the growth in compact half-mass radius since $z=2$ ($x$-axis) against the ratio between the overall and compact half-mass radii ($y$-axis) for each $z=0$ descendant.  The $x$-axis quantity is identical to the ratio between the value of the red lines in the left-hand panel at $z=0$ and $z=2$, while the $y$-axis quantity is the ratio between the black and red lines at $z=0$.  The 30\% growth in the compact half-mass radius from the changing softening length, stellar mass loss, and expulsion of gas between $z=2$ and $z=0$ that we estimate in the Appendix is marked by the vertical dotted line.   $x$-values significantly higher than this indicate some disruption of the compact core, like the mergers that caused jumps in the compact half-mass radius in the first and second examples on the left.  All of the ``consumed" galaxies, whose stellar contents have been scattered far and wide, are therefore located on the far right of this panel.  A high $y$-value indicates the presence of significant ex-situ stellar mass at large radii, so that the ``core"-type descendants separate from the ``undisturbed" descendants despite having similar $x$-values in many cases. In general, the growth in compact half-mass radius is much smaller than that of the overall half-mass radius.  This implies that a majority of the size growth is due to the addition of stellar mass in the outer parts of the galaxies.  Galaxies which gained little stellar mass typically grow by about 50\% or less in compact half-mass radius, while those which experienced some merger events are slightly more perturbed, and those which experienced majorly disruptive mergers grow even more.  

\subsection{Implications for the number density of compact galaxies}
\label{ssec:numberdensities}

As we have seen in the previous subsection, nearly all of the compact galaxies grow in size between $z=2$ and $z=0$ even if they experience little growth in mass.  Of the original 35 compact galaxies, only two remain below the compactness threshold at $z=0$.  Shifting the threshold up slightly to account for the 15\% change in half-mass radius from numerical effects yields one additional galaxy, for a final ``survival fraction" of 9\% for compact galaxies from $z=2$ to the present day.  (And of these three, one is an extreme case which dives through a massive halo, is stripped of its stellar envelope along the way, and is just escaping out the other side at $z=0$, having managed to {\it decrease} in both size and mass!)  

The number density of compact galaxies depends not only on the rate at which galaxies {\it leave} the compact population, but also the rate at which they {\it enter}.  Our previous study \citep{Wellons2015} found mechanisms for producing compact galaxies which are most efficient in the high-redshift universe when cold, dense gas is abundant.  As time increases and the universe becomes more rarefied, we expect that the production rate of compact galaxies should also decrease.  In Section \ref{ssec:cores}, we identified all the galaxies in the simulation which had a massive ($< 10^{10}$ \Msun), compact progenitor at any redshift.  For each of these galaxies, we have measured the median redshift of the compact phase as well as its total duration.  The median redshifts of the compact phases are broadly distributed from $z=1-3$, falling off at higher and lower $z$ and implying that the compact formation mechanisms indeed no longer operate as efficiently at low redshift.  Half of the compact phases lasted less than 2 Gyr, and only 10\% were longer than 5 Gyr, indicating that we should expect few massive, compact galaxies to remain in the local universe.

Both of these exercises (choosing compact galaxies at $z=2$ and tracing them forward, and finding ``ex-compact" galaxies at $z=0$ and examining the time and duration of their compact phases) indicate that the absolute number density of high-mass compact galaxies should decrease at low redshift.  Physically, this is a reflection of the growing dominance of dissipationless processes (e.g. dry mergers) at late times over processes such as disk instability or wet mergers, more common at high redshift, which provide a way to dissipate angular momentum and thus produce more compact stellar populations.

\subsection{Contextualizing the results}
\label{ssec:generalize}

Studies of compact massive high-redshift galaxies adopt varying initial selection methods, including differences in the compactness criterion (e.g. the slope and normalization of the power-law cut across the mass-radius plane), minimum stellar mass, quiescence (e.g. cuts in sSFR, UVJ, or morphology), size estimation (e.g. major-axis vs. circularized half-light radius, bulge-disk decomposition vs. single-Sersic fits), and parent surveys (e.g. SDSS, CANDELS, COSMOS).
These differences can make direct comparisons between observational surveys that infer the the mass or size evolution of massive compact galaxies difficult.
In this paper, we have primarily focused on the {\it physical} mass and size evolution of compact massive galaxies as realized in a cosmological simulation, which introduces an additional level of complexity when making comparisons.
In this subsection, we discuss some of the specific definitions used in this paper and their impact on our conclusions, which may help the reader to place these results in the context of other work, particularly observational work.

One question which plagues both simulators and observers is, what is the most meaningful way to talk about size?  
In this paper we have opted to use the stellar half-mass radius -- unambiguously defined as ``the radius of the sphere containing half of the stellar mass which is gravitationally bound to this object" -- as our fiducial measurement of size.
Using the stellar half-mass radius as a proxy for galaxy size has the disadvantages that it is not a directly observable quantity, and that it can be perturbed by the addition of stellar mass at some larger radius when the density profile is steep.
However, the half-mass radius has clear physical meaning, can be easily and robustly calculated directly from {\sc SUBFIND} data, and is related to the circularized half-light radius (modulo variations in the mass-to-light ratios for differently-aged stellar populations).  
In principle we could have adopted galaxy size estimates by fitting profiles to the stellar light distributions of every galaxy in the simulation and measuring  an effective or scale radius~\citep[e.g., as done in][]{Wellons2015}.  
However, given the complexity and variety of the simulation data the resulting radii would likely be sensitive to several details of the fitting procedure and would likely not yield characteristically different trends from those found and presented in this paper.
Nevertheless, it could be worthwhile to consider in the future a more even-handed comparison of galaxy size evolution for compact massive galaxies.

Another choice we have made in the above analysis is a particular selection of compactness threshold value, $\log(\Sigma_{1.5}) > 10.5$.  We mentioned in Section \ref{sec:general} that this selection is more severe than that employed by e.g. \citet{Barro2013}.  How sensitive are the results to this particular choice of ``compactness"?  To explore this question, we repeated the analysis with a more comparable threshold of $\log(\Sigma_{1.5}) > 10.1$, which yielded a sample of 103 galaxies in the same $1-3 \times 10^{11}$ \Msun~mass range.  These galaxies are more likely to be star-forming than the original sample of 35 (only 42\% of them have specific star formation rates lower than $3 \times 10^{-10}$/yr, as opposed to 66\% in the original sample), so more of them quickly leave the compact regime due to star formation at large radii.  When this population is traced forward, we find that they are more likely than the original sample to have descendants in the ``mixed" or ``consumed" groups, since they are larger (less tightly bound) and hence easier to disrupt via merger.  Overall, however, we find that the same trends hold as have been described above, and therefore it seems that the qualitative conclusions drawn about the diversity of growth paths for massive compact galaxies is not highly sensitive to how they are initially selected.

A more implicit set of assumptions adopted in this paper is the physical/feedback model employed by the Illustris simulation.
The masses and sizes of the simulated galaxy populations examined in this paper are specifically subject to influence from the finite mass resolution and spatial softening, pressurization of the ISM, and AGN feedback model.
We discuss in some detail the potential effects of gravitational softening (and its redshift evolution) in the Appendix of this paper.
The gravitational softening and ISM pressurization adopted in this paper can potentially prevent galaxies from fully collapsing during dissipational events such as gas-rich galaxy mergers.
In general, both gravitational softening and our method of ISM pressurization risk overestimating the sizes of galaxies and it is therefore possible that some galaxies that would at one time be compact would be missed.
Thus, our conclusions about the fractions of galaxies which pass through a compact phase might best be viewed as a lower limit.
However, the basic convergence study presented in~\citet{Wellons2015} indicated that the sizes of galaxies are well-converged at the $z=2$ gravitational softening level, so the results should not dramatically change.
Our ability to further investigate the impact of these simulation traits on our results is limited since we cannot easily re-run the Illustris simulation volume.
Reconsidering the conclusions presented in this paper with other simulations~\citep[e.g. EAGLE,][]{Schaye2015, Crain2015} or with zoom-in re-simulations of the specific objects studied in this paper or objects selected from other large full-volume simulations may provide additional context for these results in the future.

Finally, we would like to emphasize that the results herein are primarily focused on very massive galaxies with stellar masses of at least $10^{11}$ \Msun ~at $z=2$, and that caution is warranted when comparing to results in other mass regimes where different physical processes may govern the evolution.

\section{Summary and Conclusions}
\label{sec:conclusion}

Herein, we have identified a population of 35 massive ($1-3 \times 10^{11}$ \Msun) galaxies at $z=2$ in the cosmological hydrodynamical simulation Illustris which satisfy the compactness criterion $\log (\Sigma_{1.5}) > 10.5$.  We have traced these massive, compact galaxies forward in time through the simulation to their $z=0$ descendants at the present day.  Our most important findings are summarized below:

\begin{itemize}
\item Though the compact galaxies all originated at $z=2$ with the same stellar mass to within a factor of 3, by $z=0$ their stellar masses differ up to a factor of 20.  Similarly, their DM halo masses, originally within a factor of 5, spread to a factor of 40.  Rather than remaining tightly grouped, the descendants' stellar masses are dispersed throughout the entire massive galaxy population at $z=0$.
\item The compact galaxies can undertake a variety of evolutionary paths to $z=0$.  Approximately half survive as the core of a more massive descendant, about 15\% are consumed and destroyed in a merger with a much more massive galaxy, about a third are generally undisturbed by mergers, accretion, or star formation activity, and the remaining few are thoroughly mixed by major mergers.  These percentages are a weak function of initial stellar mass, with less massive galaxies more likely to be consumed and more massive galaxies more likely to be cores.
\item Very few (about 10\%) of the galaxies from the $z=2$ compact sample still satisfy our compactness criterion at $z=0$, suggesting a decrease in the number density of massive, compact galaxies at low redshift in the absence of any replenishing compact formation mechanism.
\item A compact galaxy's $z=2$ environment has some predictive power for its final evolutionary outcome.  Compact galaxies in denser environments (with the exception of satellite galaxies) are more likely to undergo mergers and gain stellar mass at larger radii, losing their compact identity in the process.  Those which are isolated at $z=2$ and experience little merger activity are also preferentially isolated at $z=0$.  Thus we predict that massive, compact relics in the local universe should predominantly live in underdense environments or as satellites in larger groups.
\item A galaxy's $z=0$ stellar mass is related to the compactness of its progenitors, such that more massive galaxies are more likely to have once been compact.  However, this trend possesses significant dispersion (partially driven by a spread in formation times) and it is {\it not} the case that the most massive galaxies at $z=0$ can be directly linked to massive compact galaxies at $z=2$.  While more than half the galaxies with stellar mass $> 10^{12}$ \Msun ~at $z=0$ have a direct compact progenitor at $z \approx 1-3$, only 15\% have a compact progenitor at exactly $z=2$.
\item The dominant source of size growth for the compact galaxies is the addition of ex-situ stellar mass at larger radii through mergers and accretion into the stellar halo, followed by disruptive major mergers, with smaller contributions from ongoing or renewed star formation, adiabatic expansion due to mass loss, and/or numerical effects.
\end{itemize}

\section*{Acknowledgments}

SW is supported by the National Science Foundation Graduate Research Fellowship under grant number DGE1144152.  PT acknowledges support from NASA ATP Grant NNX14AH35G.  CPM acknowledges support from NASA grant NNX11AI97G and NSF grant AST-1411945.  AP acknowledges support from the HST grant HST-AR-13897.  SG acknowledges support provided by NASA through Hubble Fellowship grant HST-HF2-51341.001-A awarded by the STScI, which is operated by the Association of Universities for Research in Astronomy, Inc., for NASA, under contract NAS5-26555.  LH acknowledges support from NASA grant NNX12AC67G and NSF grant AST-1312095. 

\appendix

\section{``Adiabatic expansion" from a changing gravitational softening length}
\label{app:softening}

\begin{figure*}
  \centering
  \includegraphics[width=2.1\columnwidth]{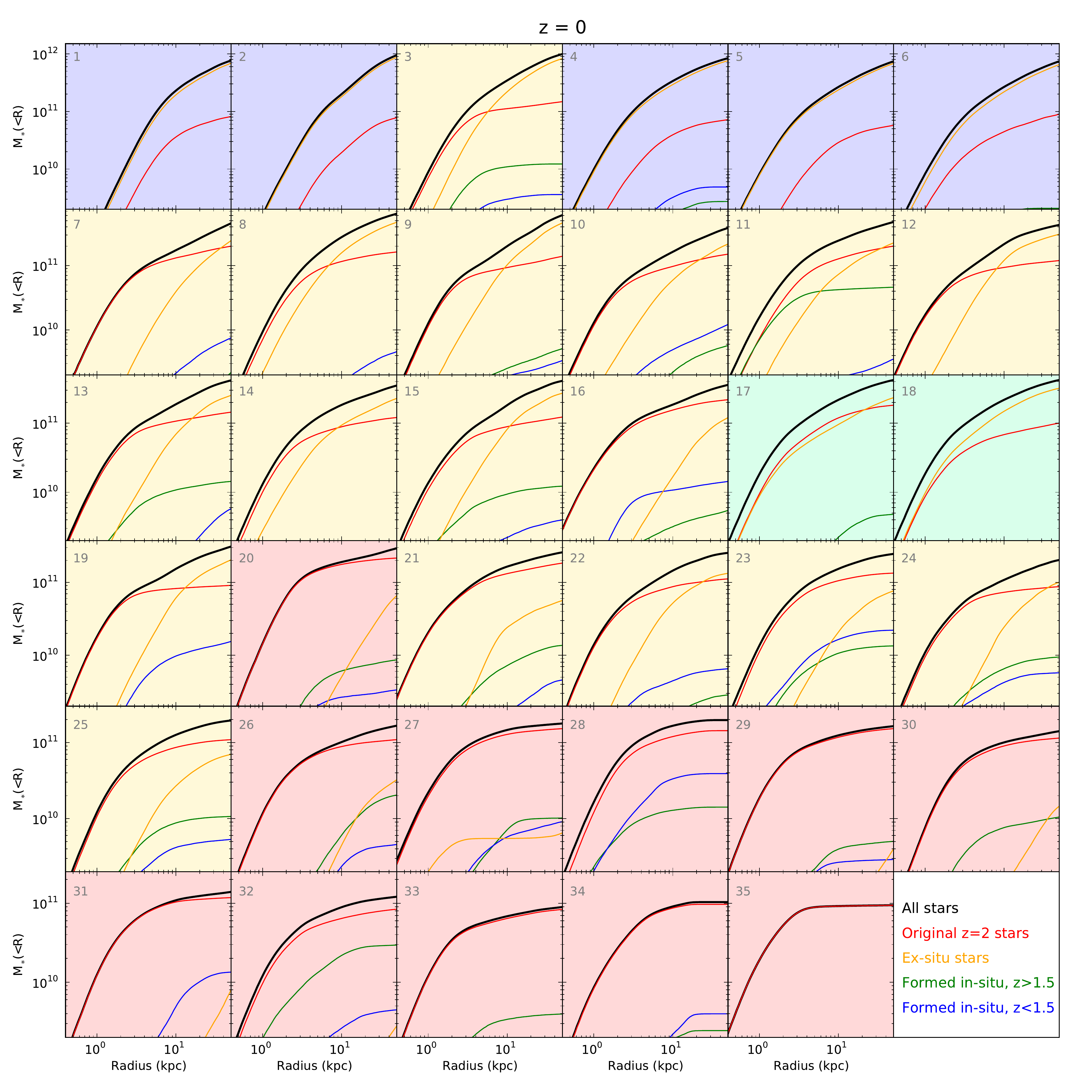}
  \caption{Cumulative stellar mass profiles for the descendants of the compact galaxies at $z=0$.  Galaxies are ordered by decreasing stellar mass, as in Figure \ref{fig:imagegrid}.  The total profiles are shown in black, and the colored lines separate the stars into {\it (red)} the original stars which were present in the $z=2$ compact progenitor,  {\it (orange)} ex-situ stars acquired after $z=2$,  {\it (green)} stars formed in-situ between $z=2$ and $z=1.5$, and {\it (blue)} stars formed in-situ after $z=1.5$.  Background shading indicates the descendant type as in Fig. \ref{fig:categories}.}
  \label{fig:profgridz0}
\end{figure*}

To estimate the effect of a changing gravitational softening length $\epsilon$ on the matter distributions of galaxies which are undisturbed between $z=2-0$, we mimic the process of adiabatic expansion that is invoked as a response to mass loss from the center of the potential on timescales longer than a dynamical time, $\tau_{\rm ej} > \tau_{\rm dyn}$.  Expansion occurs in such a way as to conserve particles' specific angular momentum $rv(r)$, which is an adiabatic invariant.  If the distribution is spherically symmetric and particles move on circular orbits, $r \propto M(<r)^{-1}$.  Changing the gravitational softening length will have much the same effect -- not by physically removing mass, but by changing how much of it an external particle ``feels."  Given a distribution of mass $\rho(r)$ and the scale $\epsilon$ on which gravity is softened, we can define an ``effective mass enclosed" $M_{\rm eff}(r, \epsilon)$ which decreases with increasing $\epsilon$.

Consider the case where gravitational forces are Plummer-softened, so that the gravitational force per unit mass $dF$ exerted by a parcel of mass $dm$ on a test particle is
$$ dF = \frac{G~dm~r_g }{(r_g^2 + \epsilon^2)^{3/2}} $$
where $r_g$ is the distance between the mass and the test particle, and $\epsilon$ is the gravitational softening length.  (Illustris employs a slightly different spline potential (see \citet{Hernquist1989, Springel2010}) but the Plummer potential is much more analytically pliable and is similar enough for our purposes here.)  If the test particle lies at a distance $R$ from a spherically-symmetric distribution of mass with density profile $\rho(r)$, we can integrate over a series of concentric shells in $r$ to determine the total gravitational force.  Each shell consists of a series of rings around the line connecting the test particle and the center, so that on a given ring (described by the angle $\theta$ from the connecting line), all points are the same distance $r_g$ from the test particle and the same distance $r$ from the center and hence exert the same force $dF \cos \phi$ along the connecting line.  By the law of cosines
$$ r_g^2 = r^2 + R^2 - 2 r R \cos \theta $$
and by the law of sines
$$\cos \phi = \sqrt{1 - \left( \frac{r}{r_g} \right)^2 \sin^2 \theta} $$
so that the gravitational force integrated over the entire sphere interior to $R$ is
$$ F_r  = \int_0^R \int_0^\pi \frac{G \rho(r) 2 \pi r^2 \sin \theta~r_g }{(r_g^2 + \epsilon^2)^{3/2}} \frac{(R - r \cos \theta)}{\sqrt{r^2 + R^2 - 2 r R \cos \theta}} d\theta dr $$
The integral over $\theta$ has an analytical solution
\begin{align*}
F_r  &= \int_0^R \frac{G \rho(r) 4 \pi r^2}{R^2} \left[ \frac{1+\epsilon^2/r^2 + R/r}{2 \sqrt{\epsilon^2/r^2 + (R/r + 1)^2}} \right. \\
 & \hspace{28pt} - \left. \frac{1+\epsilon^2/r^2 - R/r}{2 \sqrt{\epsilon^2/r^2 + (R/r - 1)^2}} \right] dr \\
 &\equiv \frac{G}{R^2} \int_0^R 4 \pi r^2 \rho(r) \beta(r, R, \epsilon) dr \\
 &\equiv \frac{G M_{\rm eff}(R, \epsilon)}{R^2}
\end{align*}
where the ``density correction function" $\beta(r, R, \epsilon)$ is the ratio between the actual mass {\it present} in a shell at radius $r$ and the mass {\it felt} by a test particle at $R$ if gravity is Plummer-softened on scales of $\epsilon$.

The form of $M_{\rm eff}(R,\epsilon)$ will vary with the density profile $\rho(r)$ and is in general not analytically solvable, so it must be evaluated numerically.  We can directly apply the above to any arbitrary mass profile, including those we recover from the simulation.  In practice, we would like to answer the following question: If a galaxy has a stellar mass profile $\rho(r)$ when the softening length is $\epsilon_1$, what will it be after the softening length is (slowly) changed to $\epsilon_2$?

\begin{figure*}
  \centering
  \includegraphics[width=2.1\columnwidth]{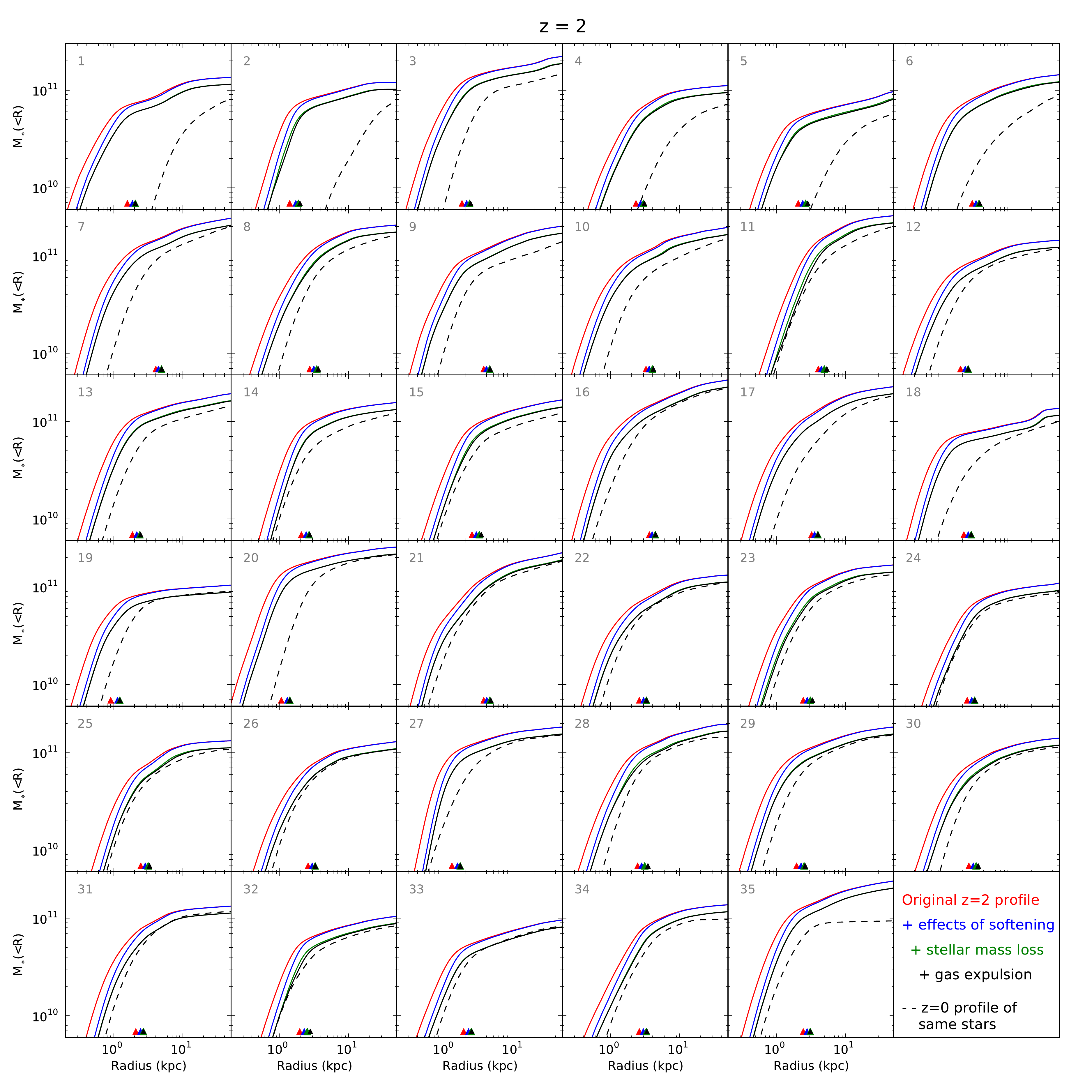}
  \caption{Cumulative stellar mass profiles of compact galaxies at $z=2$, arranged in the same manner as in Fig. \ref{fig:profgridz0}.  In each panel, the red line shows the stellar mass profile of the $z=2$ compact galaxy.  Each of the other solid lines show the cumulative effect on that original profile when {\it (blue)} the stellar and dark matter gravitational softening lengths are changed to their $z=0$ values, {\it (green)} the stars lose 15\% of their mass, and {\it (black)} any gas in the galaxy is expelled, assuming that the adiabatic invariant $rM_{\rm eff}(r, \epsilon)$ is preserved.  Triangles at the bottom of each panel mark the half-mass radii corresponding to these profiles.  On average, changing the softening length increases the half-mass radius by about 15\%, stellar mass loss then increases it by 12\%, and gas expulsion by another 3\%.  The dashed black line in each panel shows the actual stellar profile of these stars within their descendant at $z=0$.  In several cases, the dashed and solid black lines are very similar, showing that the aforementioned effects are responsible for the change in the galaxy's shape.  In most cases, however, the compact galaxy is disturbed in some way (through e.g. mergers).}
  \label{fig:profgridsoft}
\end{figure*}

To relax the profile to a new softening length, we first measure the adiabatic invariant $rM_{\rm eff}(r, \epsilon_1)$ at shells of $r_i$ in the initial profile, as well as recording the masses $m_{i-1/2}$ between the shells.  We then change the softening length to $\epsilon_2$, remeasure $r_i M_{\rm eff}(r_i, \epsilon_2)$ and compare it against the original value.  Starting with the innermost shell, we iteratively change $r_i$ until the adiabatic invariant is within some error tolerance of its initial value (placing $m_{i-1/2}$ halfway between the shells' new locations), then move on to the next shell and work our way out.

The results of performing this exercise on the compact galaxies' $z=2$ profiles are shown in Figure \ref{fig:profgridsoft}.  Each galaxy actually has two components with differing $\epsilon_2$: the (dominant) stellar component which reaches its maximum softening length of 0.7 kpc at $z=1$, and the dark matter component which continues to soften to 1.4 kpc at $z=0$.  Here we show the combined effect of both softenings on the stellar distribution, with the original $z=2$ profiles in red  and the softened profiles in blue.  On average, the changing softening length alone results in a change in half-mass radius of about $15\%$ for these galaxies.

In addition to this numerical ``mass loss," the galaxy will also physically lose mass as old stars evolve off the main sequence and explode as supernovae, and gas is expelled from the galactic center in outflows caused by stellar and black hole feedback.  These processes will also cause the stellar system to adiabatically expand.  In the green line in each panel, we repeat the above exercise with the additional step of decreasing the stellar mass in each shell by 15\% (the typical amount lost due to stellar evolution between $z=2$ and $z=0$ for this sample, see Figure \ref{fig:deltaM})  The stellar mass loss causes an additional inflation of the half-mass radius of about 12\%.  (Note that if there were no expansion in response to the mass loss, the half-mass radius would, by definition, not change, since the loss is applied equally at all radii.)  Finally, in the black line we remove any gas that was present at $z=2$, since these systems tend to be gas-poor at $z=0$, and the radius increases slightly by another 3\%.  

For comparison, the {\it actual} profiles of the same stars at $z=0$ are shown with dashed black lines.  In many cases (e.g. galaxies 21-26 and 31-34), the profile constructed by adiabatic relaxation under these few simple assumptions does a reasonably good job of reproducing the actual profile.  These cases correspond to evolutionary tracks where the compact core was not much perturbed, which includes both ``undisturbed" galaxies as well as ``core" galaxies.  Significant deviation from the estimated profile suggests events which were not taken into account in this simple model, such as major mergers.

\label{lastpage}

\end{document}